# Cooperative and Inhibitory Ion Transport in Functionalized Angstrom-Scale Two-Dimensional Channels

Mingzhan Wang[1,#], Qinsi Xiong[2,#], Gangbin Yan[1], Yu Han[1], Xiaolin Yue[1], Zhiheng Lyu[3], Zhen Li[4], Leeann Sun[1], Eli Hoenig[1], Kangli Xu[1], Nicholas H. C. Lewis[5], Kenneth M. Merz, Jr.[4], Qian Chen[3], George C. Schatz[2,*], Chong Liu[1,*]

1. Pritzker School of Molecular Engineering, University of Chicago, Chicago, IL 60637, USA.
2. Department of Chemistry, Northwestern University, Evanston, Illinois 60208, USA.
3. Department of Materials Science and Engineering, University of Illinois at Urbana-Champaign, Urbana, IL 61801, USA.
4. Department of Chemistry, Michigan State University, East Lansing, Michigan 48824, USA.
5. Department of Chemistry, Institute for Biophysical Dynamics, and James Franck Institute, University of Chicago, Chicago, IL 60637, USA.

\# equal contribution

Correspondence to: g-schatz@northwestern.edu and chongliu@uchicago.edu

**Significant success has been achieved in fabricating angstrom-scale artificial solid ionic channels aiming to replicate the biological ion channels (BICs). Besides high selectivity, BICs also exhibit sophisticated ion gating and interplay. However, such behavior and functionality are seldomly recreated in the artificial counterparts due to the insufficient understanding of the molecular origin. Here we report cooperative and inhibitory ion transport in angstrom-scale acetate functionalized MoS$_2$ two-dimensional channels. For cooperative ion transport, the permeability of K$^+$ is doubled in the presence of only 1% Pb$^{2+}$ (versus K$^+$ by molarity), while the permeability of Pb$^{2+}$ is independent of K$^+$. Molecular dynamics simulations reveal complex interplay among K$^+$, Pb$^{2+}$, and the anions in governing the cooperativity that Pb$^{2+}$ ions capture and slow down the anions via long-range interaction, which leads to the synchronization of Cl$^-$ with K$^+$ to transport as ion pairs with reduced interaction with the channel surface. For inhibitory ion transport, divalent Co$^{2+}$ (or Ba$^{2+}$) and Pb$^{2+}$ can replace each other in the confined channel and compete for the limited transport cross section. Our work reveals novel ion transport phenomena in extreme confinement and highlights the potential of manipulating ion interplay in confinement for achieving advanced functionalities.**

Biological ion channels (BICs), which are abundant in cell membranes, are critical in many cellular processes [1]. The primary signature shared by BICs is their tailored selectivity filter on the angstrom (Å) scale, which regulates specific ion transport (e.g. K$^+$, Na$^+$ and Ca$^{2+}$) across the cell membrane [2]. It has long been envisioned to replicate such marvelous BICs artificially, aiming at realizing high-performance separation of target ions and ion-based functional devices [3]. In the past two decades, significant progress has been achieved in preparing devices with Å-scale confinement with many different classes of materials including porous materials [4–8], 1D nanotubes [9–11], 2D nanochannels [12–20], thanks to advances in the precision synthesis and fabrication [21]. All these materials platforms have deepened our understanding of how water, organic solvents and ion behave in extreme confinement.

If we look at the BICs, it can quickly be realized that besides the high selectivity, rich and exquisite ion interplay phenomena play out in BICs under physiological conditions [22]. For example, in the small-conductance $Ca^{2+}$-activated $K^+$ channels, i.e. the KCNN4 channel, $Ca^{2+}$ can promote the permeability of $K^+$ via binding with calmodulin (CaM) to induce a conformational change, which in turn opens the channel pore (**Fig. 1a**) [23,24]. It is also well-established that some ions, e.g. $Ba^{2+}$, can block the $K^+$ channel (**Fig. 1b**) and that the $Ba^{2+}$ blockage is sensitive to extracellular $K^+$ concentration (the so-called "lock-in effect") [25]. Similar blocking effects are also observed in $Ca^{2+}$ channels [26,27]. All these observations suggest that there is a rich ion interplay in governing the sophisticated functions of BICs. By contrast, the predominant works on artificial channels, whether porous materials [6] or 1D nanotubes [9,10] or 2D channels [13–20], are tested with a single salt to explore ion transport behavior, with few measurements using mixed salts [28,29]. This implies that the information on ion interplay in artificial channels is far from adequate [30,31], which naturally sparks the important question: How do ions interplay in artificial channels? Can ion interplay be harnessed controllably in artificial ionic channels? Motivated by these questions, we explore ion interplay in the artificial channels. We use a $MoS_2$-based 2D channel that is covalently functionalized with acetic acid as the model system to investigate ion interplay. We have discovered two kinds of ion interplay phenomena in this system, i.e. cooperative ion transport and inhibitory ion transport. For cooperative transport, the permeability of $K^+$ is doubled in the presence of only 1% mol $Pb^{2+}$, while the permeability of $Pb^{2+}$ is independent of $K^+$. Molecular dynamics (MD) simulation reveals the molecular mechanism underpinning this complex ion behaviors that mobile $Pb^{2+}$ plays a key role in slowing down the anions via long-range interactions, which influences the pairing between $K^+$ and anions in the two-dimensional channels. The correlated transport between $K^+$ and anions in the channel reduces the $K^+$ interaction with channel surface, and therefore enhances its permeability. For inhibitory ion transport, divalent $Co^{2+}$ (or $Ba^{2+}$) cannot pair with monovalent anions due to its slow diffusion and compete strongly with $Pb^{2+}$ to enter the confined channel and interact with the functional groups on the channel surface, which result in the reduced flux.

**Experimental design**

To explore the ion interplay in the artificial 2D channels, we designed the experimental procedure shown in **Fig. 1c**. We have systematically examined the permeabilities of individual ions (Cases 1 and 2 in **Fig. 1c**) and for mixed ions (Case 3 in **Fig. 1c**) and compare them accordingly. The model 2D material platform we chose is acetic acid-functionalized $MoS_2$ ($MoS_2$-COOH, **Fig. 1d**,**e**, and **Supplementary Fig. 1**) for the following two reasons: first, the covalently bonded carboxyl groups can lead to controlled hydration of the $MoS_2$-COOH membrane [32], which, for comparison, cannot be achieved using the unfunctionalized $MoS_2$ membrane [33,34]; second, our previous work has shown that carboxyl groups have strong chelating capability with metal ion [35], which will be discussed later. 2D channels are built by stacking $MoS_2$-COOH nanosheets using vacuum filtration. A cross-sectional scanning transmission electron microscopy (STEM) image clearly shows the layered structure of the restacked $MoS_2$-COOH membrane in its dry state (**Fig. 1d**). Details of functionalization method and membrane preparation are provided in the Supplementary Materials.

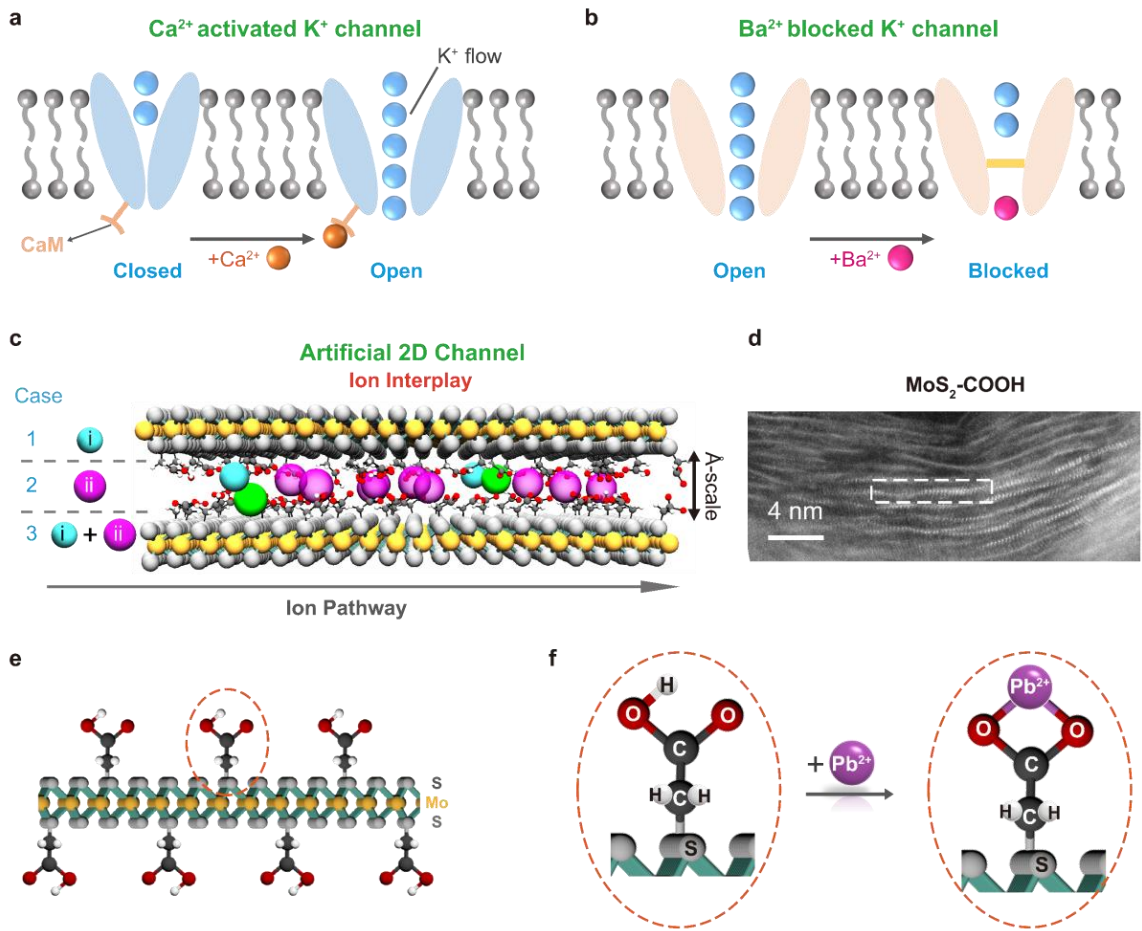

**Fig. 1 Exploring ion interplay in artificial 2D channels.** (**a**) Schematic of the working mechanism of the $Ca^{2+}$-activated $K^+$ channels, where calmodulin (CaM) can open the channel upon binding $Ca^{2+}$. (**b**) Schematic of the working mechanism of the $Ba^{2+}$-blocked $K^+$ channels. (**c**) Experimental procedure for exploring ion interplay in artificial 2D channels. i and ii indicate different ions. (**d**) Cross-section of STEM image of $MoS_2$-COOH membrane showing the 2D channels in the dry state. (**e**) Schematic of the structure of the $MoS_2$-COOH nanosheet, as boxed out in (**d**). Orange circle indicates a covalently bonded carboxyl group attached to the $MoS_2$ plane, which is illustrated in detail in (**f**, left). (**f**) Schematic showing the $Pb^{2+}$ induced deprotonation and its bidentate binding mode with the acetate functional group.

In the $MoS_2$-COOH system, we particularly focus on $Pb^{2+}$, because our previous study has revealed: i) it can induce deprotonation of the carboxyl functional group and ii) $Pb^{2+}$ uniquely adopts a bidentate binding mode with the carboxylate functional group (**Fig. 1f**) and locks the interlayer spacing, as confirmed by FTIR measurements and DFT calculations [35]. High-angle annular dark-field (HAADF) STEM image clearly shows the presence of single-site $Pb^{2+}$ in the $MoS_2$-COOH membrane treated by $Pb^{2+}$ solution **(Supplementary Fig. 2)**. We focus our study on binary ion interplay between $Pb^{2+}$ and monovalent ions ($K^+/Li^+$) or between $Pb^{2+}$ and divalent ions ($Co^{2+}$ and $Ba^{2+}$). We conducted permeation transport tests using an H-type cell, where one chamber is filled with feed salt solution and the other filled with an equal volume of deionized water, with the membranes mounted in between **(Supplementary Fig. 3)**. The transmembrane permeabilities of $Pb^{2+}$ and other ions ($K^+$, $Li^+$, $Co^{2+}$ and $Ba^{2+}$, with ion parameters listed in **Supplementary Table 1**) were measured using single ion solutions and corresponding binary mixture solutions sequentially, as mentioned in **Fig. 1c**. Other experimental details and the method for calculating permeability are detailed in the Supplementary Materials. All the salt solutions were tested with natural pH values after dissolving in deionized water without further tuning (pH values ranging from ~3.6 to ~6.6).

**Cooperative ion transport**

- **Experimental Results**

We started by examining the interplay between $K^+$ and $Pb^{2+}$ (both salts are nitrate-based). The results are summarized in **Fig. 2a**, with more detailed results in **Supplementary Fig. 4** and **Supplementary Table 2.** Each bar of **Fig. 2a** refers to the ion flux into the permeate observed in a sequence of experiments in which the membrane is exposed to different feed solutions indicated at the bottom. Thus, the first bar in **Fig. 2a** refers to $K^+$ only. Then the feed is switched to $Pb^{2+}$ only with the same membrane for the second bar, and followed by $K^+$ only in the third bar and mixtures of $K^+$ and $Pb^{2+}$ in the subsequent bars. These results show that the permeability of $K^+$ (third bar) decreases by ~20% after the $Pb^{2+}$ test when compared to the initial permeation test of $K^+$ (first bar). More interestingly, with a tiny fraction of $Pb^{2+}$ (1% by mol $K^+$) added into the feed (4$^{th}$ bar), the permeability of $K^+$ is doubled. After that the permeability of $K^+$ continues to rise with the addition of more mobile $Pb^{2+}$ in the feed, and reaches a peak value ~ 350% when the $Pb^{2+}/K^+$ ratio is 25% (8$^{th}$ bar, as also shown in **Fig. 2b**). Further increase of the $Pb^{2+}/K^+$ ratio beyond 25% leads to a slight decrease of $K^+$ permeation. In contrast to this, the permeability of $Pb^{2+}$ only scales with the feed concentration of $Pb^{2+}$ (**Fig. 2b**), regardless of the presence of $K^+$ or not (**Supplementary Fig. 5**). Similarly enhanced transport is also observed in the $Li^+/Pb^{2+}$ pair tests (**Supplementary Fig. 6**, nitrate anions) and $K^+/Pb^{2+}$ pair tests with chloride anions instead of nitrate anions (**Supplementary Fig. 7**) in the $MoS_2$-COOH 2D channel. Our results are also contrasted with previous works reporting suppressed transport of a monovalent cation by a divalent cation in nanotube [28] and nanopore tests [29]. Clearly, all our results unambiguously confirm that there is clear ion interplay between $K^+$ (or $Li^+$) and $Pb^{2+}$ when they transport in the 2D $MoS_2$-COOH channels and the ion interplay is asymmetric. We term this asymmetric ion interplay as a *cooperative ion transport* effect. Additionally, the cooperative ion transport effect is absent in tests using the polytetrafluoroethylene (PTFE) membrane with large pores of ~200 nm (**Supplementary Fig. 8**), where the permeabilities of $K^+$ and $Pb^{2+}$ are totally independent of each other (**Fig. 2c** and **Supplementary Fig. 9**) and the measured permeability ratio of $K^+/Pb^{2+}$ across the PTFE membrane is ~ 1.76, close to their bulk diffusion coefficient ratio of ~ 2.07. The results indicate independent bulk ion transport properties in the large pore PTFE membrane without any ion interplay, highlighting the significance of confinement to ion interplay.

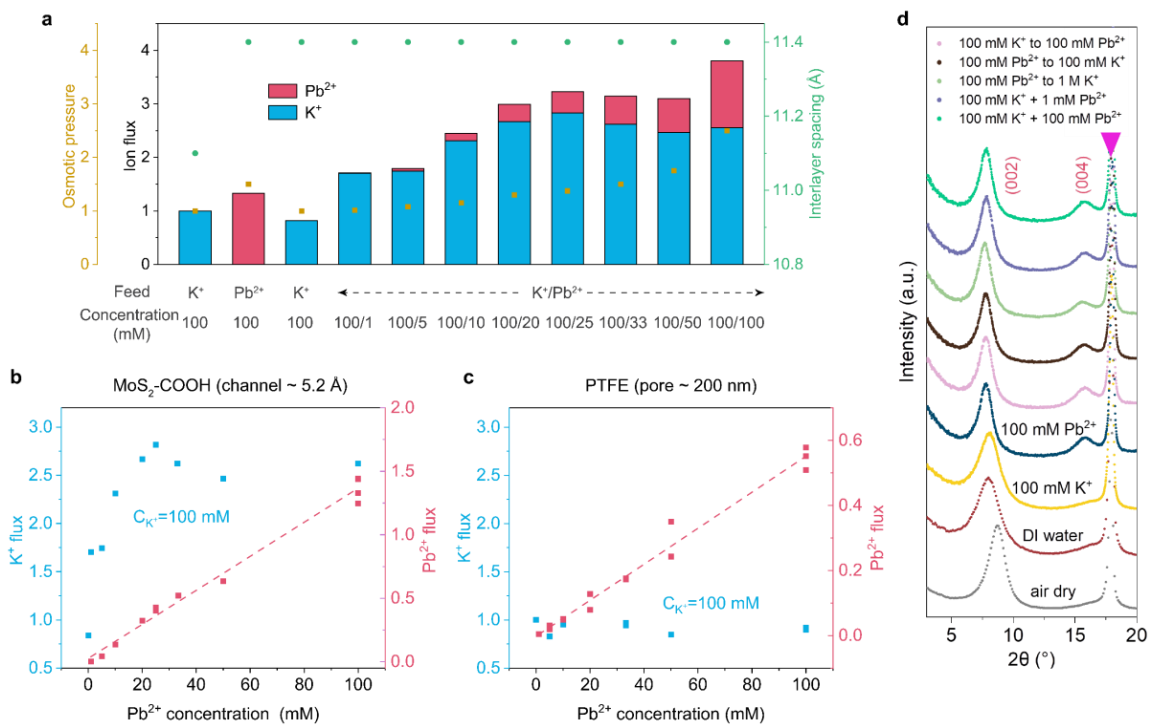

**Fig. 2 Cooperative ion transport.** (**a**) Summary of the sequential permeation test results of $K^+$, $Pb^{2+}$, and $K^+/Pb^{2+}$ mixtures across one $MoS_2$-COOH membrane. (**b**, **c**) The fluxes of $K^+$ and $Pb^{2+}$ versus concentration of $Pb^{2+}$ in the feed for the series of tests across the $MoS_2$-COOH membrane and PTFE membrane, respectively. In the panel, both the ion fluxes and osmotic pressure are normalized to values corresponding to the initial $K^+$ test. The counter ion is nitrate. (**d**) Systematic XRD measurements for the $MoS_2$-COOH membrane. The effective height of the $Pb^{2+}$-locked channel is ~ 11.4 Å-6.2 Å=5.2 Å, where 6.2 Å corresponds to the intrinsic thickness of monolayer $MoS_2$. The annotation "solution A to solution B" (in the legend) indicates that the $MoS_2$-COOH membrane was first saturated in solution A and then switched to solution B for uptake, and finally the $MoS_2$-COOH membrane was taken out of solution B for XRD measurements. The pink triangle indicates the PTFE peak.

To the best of our knowledge, this is the first such observation of cooperative ion transport effect in artificial channels in the nanofluidics studies. To understand the mechanism underlying this cooperative ion transport, we first calculated the osmotic pressure change, given that the ion permeation tests are driven by osmotic pressure. Clearly, as plotted in **Fig. 2a**, the enhanced permeation of $K^+$ cannot be solely attributed to the change in osmotic pressure, because the enhancement is well above the change of osmotic pressure, particularly in the low $Pb^{2+}/K^+$ regime, e.g. 1% mol $Pb^{2+}$ addition. We also notice that the cooperative ion transport observed bears striking resemblance to the $Ca^{2+}$-activated $K^+$ BICs, where enhanced $K^+$ transport is attributed to the pore opening due to conformational change induced by CaM upon binding $Ca^{2+}$. Inspired by the pore opening mechanism, we systematically examined the interlayer spacings of $MoS_2$-COOH channels in a series of tests for $K^+/Pb^{2+}$. However, the results show that after the single $Pb^{2+}$ test (2$^{nd}$ bar in **Fig. 2a**), the 2D $MoS_2$-COOH confinement is fixed (or locked) by $Pb^{2+}$ with an effective channel height ~ 5.2 Å (**Fig. 2d**). The locking effect of $Pb^{2+}$ is not surprising, given the dominant uptake of $Pb^{2+}$ as compared with $K^+$ in the $MoS_2$-COOH membrane (**Supplementary Fig. 10**). Apparently, this cooperative ion transport cannot be attributed to a change in confinement size either and points to an origin different from the $Ca^{2+}$-activated $K^+$ channels.

- **MD simulations and mechanism**

To elucidate the mechanism underlying our observations, we conducted all-atom MD simulations. We conducted a sequence of permeation tests in the same order as the experiments, and the simulated layer spacings were also consistent with the experiments. Specifically, each bar in **Fig. 3B** represents the ion flux into the permeate as observed in a series of calculation tests, with the specific ion numbers for the feed solutions indicated at the bottom of the figure. The first bar represents the flux with a feed solution containing only K$^+$ ions and no Pb$^{2+}$ ions in the membrane. The second bar represents the flux after switching the feed solution to contain only Pb$^{2+}$ ions, using the same membrane. The third bar represents the flux with a feed solution containing only K$^+$ ions. From the third bar onward, the membrane contains Pb$^{2+}$ ions to neutralize the deprotonated -COO$^-$ in the channel. The subsequent bars represent the feed solution with a mixture of K$^+$ and Pb$^{2+}$ ions. The parameters are outlined in the Supplementary Materials (**Supplementary Tables 3** and **4)** and additional details regarding the permeation tests in the MD simulations can be found in the Supplementary Materials and **Supplementary Table 5**. We simulated ion transport from the feed into the membrane by applying high pressure to the piston wall (**Fig. 3a**), with the applied pressure changes (**Fig. 3b**) consistent with the osmotic pressure changes shown in **Fig. 2a**. Ion fluxes for a series of permeation tests are presented in **Fig. 3b** and **Supplementary Fig. 14**. The fluxes of K$^+$ and Pb$^{2+}$ in our simulations exhibit a trend in good agreement with the experimental results. Particularly with one Pb$^{2+}$ added into the feed (Test 4), the flux of K$^+$ is enhanced by ~ two-fold compared to that in Test 3 without mobile Pb$^{2+}$ in the feed. Note that limited by the simulation scale, the observed Pb$^{2+}$ flux here includes Pb$^{2+}$ from the membrane, and therefore is much larger than what is observed in experiments. Nevertheless, our model provides a reliable qualitative explanation for the increase in Pb$^{2+}$ flux.

To explain the flux changes, we first calculated the diffusion coefficients of K$^+$/Pb$^{2+}$/Cl$^-$ (**Fig. 3c**). The diffusion coefficients of Pb$^{2+}$ in both the feed (**Fig. 3c**) and the membrane (**Supplementary Fig. 15**) increase with mobile Pb$^{2+}$ concentration in Test 4 and thereafter, which is consistent with experiment (**Fig. 2b**). Therefore, we attribute the increased flux of Pb$^{2+}$ to the effect of applied pressure (equivalent of osmotic pressure). Evidently, the diffusion coefficients of $K^+/Cl^-$ in Test 1 are markedly higher than those in Test 3. This is attributed to the fact that the membrane locked by $Pb^{2+}$ in Test 3 has deprotonated -COO⁻ groups, which attract $K^+$ and thus reduce its flux. The flux of $K^+$ increases from Test 3 to Test 7 and starts to decrease thereafter, similar to the experimental results. The increase in applied pressure can also explain the increase of $K^+$ flux from Test 5 to Test 7. After Test 7, as $Pb^{2+}$ diffuses more rapidly, it escapes trapped sites, allowing more vacant -COO⁻ groups to capture $K^+$, resulting in decreased $K^+$ flux (**Supplementary Fig. 16**). However, the applied pressure change cannot explain the difference between Test 3 and Test 4, as they have the same pressure with only 1 $Pb^{2+}$ / 2 $Cl^-$ ion difference, whereas the $K^+$ fluxes differ by a factor of two.

To distinguish whether the two-fold increase of $K^+$ permeability in Test 4 was caused by $Pb^{2+}$ or $Cl^-$, we replaced $1PbCl_2$ with 2KCl in the feed (Test 9, **Supplementary Table 5**). The result shows that increasing the concentration of KCl does increase the flux of $K^+$, but only by 30-40% (**Supplementary Fig. 17**), well below the two-fold increase observed in Test 4. Therefore, the two-fold increase in $K^+$ flux observed in Test 4 is mostly caused by the additional $Pb^{2+}$. As shown in **Fig. 3c**, due to the very low diffusion coefficient of mobile $Pb^{2+}$, its entry into the membrane is much slower than that of other ions. Consequently, we hypothesize that the main place where mobile $Pb^{2+}$ plays a role is in the feed. To verify this, we further calculated the radial distribution function (RDF) of each $Cl^-$ (relative to $Pb^{2+}$) before entering the membrane (**Supplementary Fig. 18, a-c**), as well as the diffusion coefficients of these $Cl^-$ (**Supplementary Fig. 18, d-f**). We find that approximately 15-20% of $Cl^-$ ions are close to $Pb^{2+}$ ions (within 10 Å), and the diffusion coefficient of these $Cl^-$ in the feed is only ~1/3 that of $Cl^-$ ions not within 10 Å (**Fig. 3d**). Meanwhile, the diffusion coefficients of water molecules near mobile $Pb^{2+}$ are reduced by 30%-50% compared to those far away from $Pb^{2+}$ (**Supplementary Fig. 19**). This result is consistent with previous research on $Pb^{2+}$ and other divalent ions [36–38], with the key difference being that the previous study focused on the interactions between $Pb^{2+}$ ions and water molecules in the first or second solvation shell, whereas our investigation suggests that interactions beyond the second solvation shell (out to 10 Å from $Pb^{2+}$) are important as well. Moreover, we find that $K^+$ in the $MoS_2$-COOH channel is easily trapped by acetate, while the presence of $Cl^-$ facilitates the movement of $K^+$. $K^+$ tends to either form $K^+/Cl^-$ ion pairs with $Cl^-$ in the membrane (**Fig. 3e** and **Supplementary Fig. 20**) or $K^+$ follows the trajectory of $Cl^-$ moving out of the channel (**Supplementary Fig. 21**).

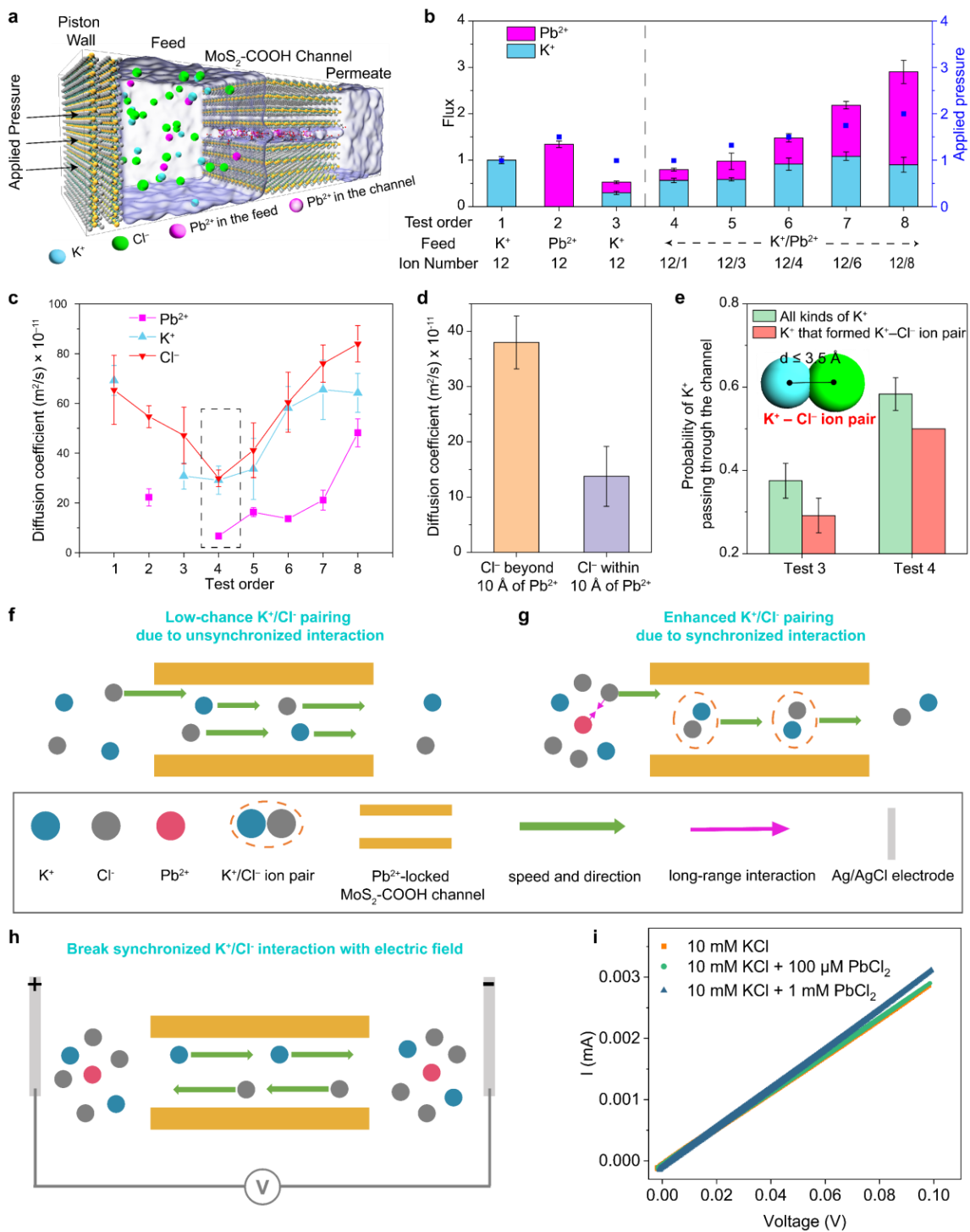

**Fig. 3 MD simulations and mechanism.** (**a**) Simulation modeling of ion transport through 2D MoS$_2$-COOH channels, which includes six extra MoS$_2$ sheets as the supporting membrane (Mo in yellow and S in gray), water (transparent ice blue), ions (in cyan, green, and magenta) and two MoS$_2$ sheets as piston wall. Note that Pb$^{2+}$ here are divided into mobile Pb$^{2+}$ in the feed and Pb$^{2+}$ in the membrane (shown in semi-transparent magenta

beads for clarity), because additional $Pb^{2+}$ is needed to neutralize the deprotonated -COO$^-$ in the channel. (**b**) MD simulation results analogous to the experimental results in **Fig. 2a** for the fluxes of $Pb^{2+}$ and $K^+$ across the MoS$_2$-COOH channel with $K^+/Pb^{2+}$ concentration ratios indicated for each test. In the panel, both the ion fluxes and applied pressure are normalized to values corresponding to the initial $K^+$ test. Note that the flux of $Pb^{2+}$ (purple bar) observed in Test 3 originates from the $Pb^{2+}$ present in the membrane. (**c**) The diffusion coefficients of $K^+$, mobile $Pb^{2+}$ and $Cl^-$ throughout the simulations. The dotted framework indicates synchronized interaction between $K^+$ and $Cl^-$. (**d**) Diffusion coefficients for $Cl^-$ beyond and within a 10 Å radius of mobile $Pb^{2+}$ before entering the membrane for the simulation in Test 4. (**e**) The probability of all kinds of $K^+$ passing through the channel (pale greenish bar) and that of $K^+$ that are knocked by $Cl^-$ to form a $K^+/Cl^-$ ion pair passing through the channel (pink bar). Inset shows a snapshot of the $K^+/Cl^-$ ion pair (within 3.5 Å) inside the channel for Test 4. (**f**) Schematic showing the permeation of $K^+$ and $Cl^-$ with poor ion pair formation in the $Pb^{2+}$-locked MoS$_2$-COOH channel in the absence of mobile $Pb^{2+}$ in the feed. (**g**) Schematic showing the permeation of $K^+$ and $Cl^-$ with enhanced ion pair formation in the $Pb^{2+}$-locked MoS$_2$-COOH channel with mobile $Pb^{2+}$ slowing down $Cl^-$ via long-range interaction. (**h**) Schematic showing ion migration under electric field. Synchronized interaction between $K^+$ and $Cl^-$, i.e. ion pair formation between $^+$ and $Cl^-$ is absent in the channel due to their opposite migration direction. (**i**) Measured I-V curves in the $Pb^{2+}$-locked MoS$_2$-COOH channel.

Combining the systematic analysis above, the cooperative transport is attributed to the complex interactions among $Pb^{2+}$, $Cl^-$, and $K^+$. In the absence of mobile $Pb^{2+}$ in the feed (Test 3 and Test 9), the larger diffusion coefficient of $Cl^-$ compared to $K^+$ (**Fig. 3c** and **Supplementary Fig. 22**) reduces the frequency of $K^+$ and $Cl^-$ collisions, resulting in less ion pair formation (**Fig. 3e,f**). The probability of ion pair formation in Test 3 is only ~55% of that in Test 4 (**Fig. 3e**). Additionally, a potential of mean force calculation shows that $Pb^{2+}$ and $Cl^-$ that are within 10 Å have relatively strong long-range interactions (**Supplementary Fig. 23**). Therefore, in the presence of mobile $Pb^{2+}$ in the feed, even $Cl^-$ beyond the second solvation shell of $Pb^{2+}$ (5-10 Å) can be captured to be slowed down (**Fig. 3d**), which leads to its synchronization of $Cl^-$ with $K^+$ to form ion pairs (**Fig. 3g**). Consequently, ion-pair formation opens up pathways for $K^+$ diffusion, significantly increasing the flux of $K^+$.

To verify the proposed mechanism, we measured the electric field driven transport of $K^+$ and $Cl^-$ using KCl solution (with and without mobile $Pb^{2+}$) in the $Pb^{2+}$-locked $MoS_2$-COOH membrane. In the electric field driven transport tests, the movement of $K^+$ and $Cl^-$ are in opposite directions in the channel (**Fig. 3h** and **Supplementary Fig. 24,** more details in the Supplementary Materials), as opposed to their movement in the same direction in the osmotic pressure driven permeation test (**Fig. 3f,g**). In other words, the synchronization-induced enhanced pairing between $K^+$ and $Cl^-$ in the channel is absent under the electric field. The results show that 1% mol $Pb^{2+}$ addition almost makes no change to the conductance (i.e. ion flux) and 10% mol $Pb^{2+}$ addition only enhances the conductance only by ~ 8.7% (**Fig. 3i**), which is much lower than the large enhancement of $K^+$ permeability in the presence of corresponding $Pb^{2+}$ ratios in the permeation tests (~2-4 fold, **Supplementary Fig. 7** and **Fig. 2b**). These contrasting results therefore corroborate the ion pairing mechanism as we propose above.

**Inhibitory ion transport**

Besides monovalent ions, we also studied the interplay between $Pb^{2+}$ and divalent cations. Divalent ions have slower diffusion coefficients, it is important to examine whether monovalent anions can be slowed down by $Pb^{2+}$ via long-range interaction to the level that it can form ion pairs to enable cooperative ion transport for divalent cations. Given the importance of dehydration to ion transport under confinement [17,19], $Co^{2+}$ and $Ba^{2+}$ are chosen because $Ba^{2+}$ has one of the smallest hydration enthalpy (-1332 kJ/mol) among divalent ions, while $Co^{2+}$ has one of the largest (-2036 kJ/mol, **Ref.**[39] and **Supplementary Table 1**). Moreover, our previous work has revealed that both $Co^{2+}$ and $Ba^{2+}$ adopt a monodentate binding mode with the acetate functional group (**Supplementary Fig. 25**) [35]. We observed a different ion interplay between $Pb^{2+}$ and $Co^{2+}$ (or $Ba^{2+}$). The permeability of either $Co^{2+}$ or $Ba^{2+}$ decreases with increasing $Pb^{2+}$ concentration in the feed, while the permeability of $Pb^{2+}$ increases with increasing $Pb^{2+}$ concentration in the feed, but in a seemingly nonlinear way (**Fig. 4a,b**, **Supplementary Figs. 26** and **27**, **Supplementary Tables 6** and **7**). The relatively higher $Ba^{2+}$ flux as compared with $Co^{2+}$ can be attributed to the fact that $Ba^{2+}$ has a smaller hydration enthalpy than $Co^{2+}$, which means a smaller dehydration barrier for $Ba^{2+}$ than $Co^{2+}$ in the $Pb^{2+}$-locked $MoS_2$-COOH channels. Given that this is the opposite to the cooperative ion transport observed between $K^+$ and $Pb^{2+}$, we term the ion interplay between $Co^{2+}$ (or $Ba^{2+}$) and $Pb^{2+}$ as *inhibitory ion transport*. Similar to the cooperative ion transport case, inhibitory ion transport is absent in the large pore PTFE test (**Supplementary Fig. 28**).

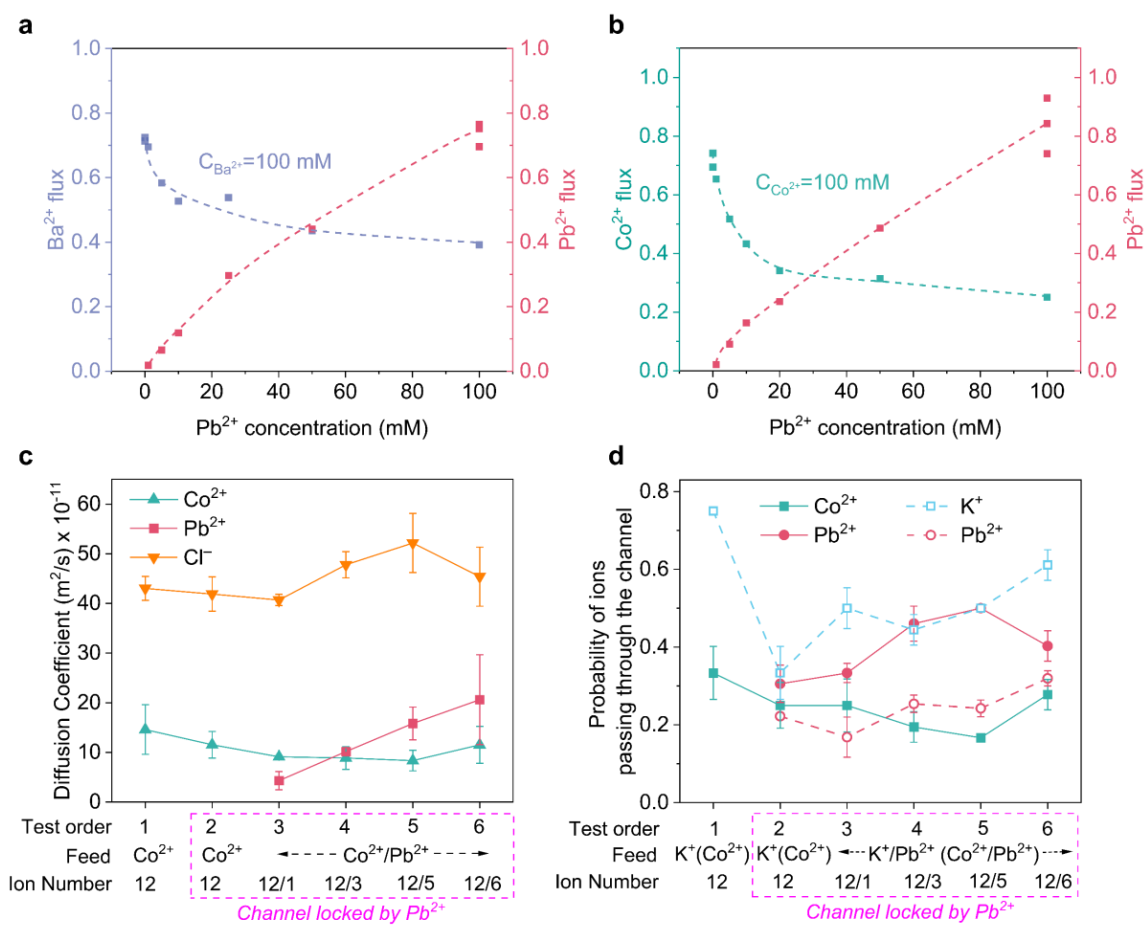

**Fig. 4 Inhibitory ion transport.** (**a**) Measured fluxes of $Ba^{2+}$ and $Pb^{2+}$ versus the concentration of $Pb^{2+}$ in the feed in the series of $Ba^{2+}/Pb^{2+}$ pair permeation tests across one $Pb^{2+}$-locked $MoS_2$-COOH membrane. (**b**) Measured fluxes of $Co^{2+}$ and $Pb^{2+}$ versus the concentration of $Pb^{2+}$ in the feed in the series of $Co^{2+}/Pb^{2+}$ pair permeation tests across one $Pb^{2+}$-locked $MoS_2$-COOH membrane. In (**a**) and (**b**), both the ion fluxes are normalized to values corresponding to the initial $Ba^{2+}$ test and $Co^{2+}$ test, respectively. (**c**) Calculated diffusion coefficients of $Co^+$, mobile $Pb^{2+}$ and $Cl^-$ throughout the MD simulations. (**d**) Probability of ions passing through the channel. Solid lines and dashed lines represent $Co^{2+}/Pb^{2+}$ and $K^+/Pb^{2+}$ tests, respectively.

To figure out the mechanism of the inhibitory ion transport, we conducted MD simulations of the $Co^{2+}/Pb^{2+}$ pair tests. Settings for the permeation tests in the MD simulations are detailed in **Supplementary Table 8**. In the simulations, we observe a reduced trend of $Co^{2+}$ flux (**Supplementary Fig. 29**) with increasing $Pb^{2+}$ concentration in the feed, consistent with the experimental results. We also investigated the ion flux into the membrane (**Supplementary Fig. 30**) and observed that the overall ion fluxes in the $Co^{2+}/Pb^{2+}$ test are lower than that in the $K^+/Pb^{2+}$ test. Specifically, due to the higher dehydration enthalpy of $Co^{2+}$, its flux into the membrane was significantly lower than that of $K^+$. Additionally, the diffusion coefficient of $Co^{2+}$ is much lower as compared with $Cl^-$ (**Fig. 4c**), suggesting a lower chance of forming ion-pairs between $Co^{2+}$ and $Cl^-$. More $Pb^{2+}$ ions pass through the channel in the $Co^{2+}/Pb^{2+}$ test while more $K^+$ ions pass through the channel in the $K^+/Pb^{2+}$ test (**Fig. 4d**). This suggests that $Co^{2+}$ can replace $Pb^{2+}$ within the membrane due to its strong binding affinity with the membrane.

As the concentration of mobile $Pb^{2+}$ increases, the flux of $Co^{2+}$ is impeded. This can be interpreted as that in Å-scale confinement, bypassing ions that block the channel is unlikely, therefore $Co^{2+}$ and $Pb^{2+}$ cannot permeate independently and have to compete for the limited transport cross section (i.e., area of the channel available for transport) in the 2D channels. Therefore the increase of $Pb^{2+}$ flux will decrease the flux of $Co^{2+}$ and vice versa.

To conclude, we have provided conclusive experimental evidence of complex ion interplay under angstrom-scale 2D confinement. Several issues were found to contribute to this interplay, including the locking of layer separation by $Pb^{2+}$ ions, the strong binding of $Pb^{2+}$ to acetate such that once the membrane has been exposed to $Pb^{2+}$, the layer spacing doesn't change to accommodate the other ion. This leads to an osmotic pressure that increases with increasing $Pb^{2+}$ concentration, which leads to increasing $K^+$ diffusion given that the acetate sites are less likely to be bound to $Pb^{2+}$. In addition, we found that cooperative transport is mediated by the anion, which more effectively forms $K^+$-anion ion pairs in the presence of $Pb^{2+}$ due to slowing of $Cl^-$ diffusion in the presence of $Pb^{2+}$. These effects lead to cooperative ion transport, which has not been reported in nanofluidic devices. Due to the Å-scale spacing in the 2D $MoS_2$-COOH channel, but inhibitory ion transport is observed, when the diffusion coefficient of cations is too slow to catch up with the anions, such as for divalent metal cations ($Co^{2+}$ and $Ba^{2+}$).

In further work, our methods can be extended to study ion interplay in channels with varied confinement size compared to $MoS_2$, with more sophisticated confinement chemistry and with more complex ion combinations than binary combinations. For example, in the 2D confinement, one direction is to study ion interplay in a channel with dynamic change in size, e.g. $Pb^{2+}/Al^{3+}$ pair in $MoS_2$-COOH channel (**Supplementary Fig. 31**). Altogether, all these efforts not only unveil novel mass transport rules in extreme confinement, but also point to the great potential of manipulating ion transport and achieving responsive ion transport based on chemical environment, which will advance the design of ion-based functional devices.

## Methods
### Chemicals:

$KNO_3$ (≥99%, Sigma-Aldrich), $Co(NO_3)_2 \cdot 6H_2O$ (98%, Sigma-Aldrich), $Pb(NO_3)_2$ (>99%, Sigma-Aldrich), $LiNO_3$ (99+%, Acros Organics), $Ba(NO_3)_2$ (99.999%, Alfa Aesar), KCl (99.0-100.5%, Alfa Aesar), $PbCl_2$ (≥99 %, Oakwood Products Inc), $MoS_2$ powders (Sigma-Aldrich) iodoacetic acid ($ICH_2COOH$, Sigma-Aldrich) were used as received without further treatments. All the reactions were done at room temperature.

## Preparation of $MoS_2$-COOH dispersion solution

First the single- or few-layer $MoS_2$ nanosheet aqueous dispersion solution was prepared through the intercalation of $MoS_2$ powders (Sigma-Aldrich) in **n-butyl lithium/hexane solution** (1.6 M) (**Caution!**), as widely reported previously. Then the acetic acid functionalization of the $MoS_2$ nanosheet was achieved through its nucleophilic reaction with iodoacetic acid ($ICH_2COOH$), as we reported in **Ref. 32** of the main text. **Our previous measurement shows that the functionalization degree is ~ 25% -COOH per $MoS_2$.**

Note that before membrane preparation, to remove possible ions contaminations, the prepared $MoS_2$-COOH solution were washed with dilute trace-metal HCl (~2-5 mM), followed by filtration and redispersion of $MoS_2$-COOH in water under sonication (~10-15 min). The "HCl washing-filtration-redispersion" cycle was repeated at least three times. Then, the redispersed $MoS_2$-COOH solution was dialyzed in dialysis tubings (Sigma-Aldrich, Part No. D9527-100FT) to remove extra HCl. Finally, repeated centrifugations (~4000 rpm/10min) were used to get the aqueous dispersion solutions of single- or few-layer $MoS_2$-COOH nanosheets (concentration 0.1-0.2 mg/ml), which were stored under 4 °C for use.

## Preparation of the $MoS_2$-COOH membrane

The $MoS_2$-COOH membrane was prepared via the vacuum assisted filtration method. The substrate used is hydrophilic polytetrafluoroethlene (PTFE) (**Supplementary Fig. 8**, pore size: 0.2 μm, purchased from Sigma-Aldrich, Part No. JGWP04700). Upon filtration, the $MoS_2$-COOH/PTFE complex membrane was transferred to clean Petri dishes for natural air drying. Note that to avoid membrane shrinkage or deformation in the drying process, several drops of DI water were put beneath the complex membrane before membrane transfer to remove air bubbles between PTFE and dishes. After the air-drying process, the prepared complex membranes were used for ion transport tests, membrane uptake tests and other characterizations.

**The sequential permeation tests across one $MoS_2$-COOH membrane**

We used a pair of H-cells to conduct the series of permeation tests of binary mixture solutions across the membranes, including the porous PTFE membrane and $MoS_2$-COOH/PTFE complex membrane. The membranes were mounted in between the H-cells by commercial Silicon paste (**Supplementary Fig. 3**). The H-cells are ready for permeation tests upon solidification of the Silicon paste. All the permeation tests were done at room temperature.

In a typical permeation test, 20 ml solution (either single salt solution or binary mixture) and 20 ml deionized water were simultaneously poured into the feed side and the permeate side of H-cells, respectively. Note that the $MoS_2$-COOH faces the feed side in the $MoS_2$-COOH/PTFE complex membrane tests. Both chambers were strongly stirred to mitigate concentration polarization. The concentration changes of every metal cation in the permeate side were monitored through periodic samplings (20-60μl per sampling subject to change according to permeation time), which were diluted by 3% $HNO_3$ (trace-metal grade) to 3-5 ml for standard ICP-MS measurements. Both chambers were tightly sealed with parafilm to avoid undesired evaporation throughout the process, except when sampling. We observed no obvious level differences developed between feed and permeate in the short test time (~ 2 hours).

The sequential permeation tests across one $MoS_2$-COOH/PTFE membrane were conducted by switching the feed solutions. **After every switching, both the permeate and the feed chambers were thoroughly washed with DI water several times to remove ions leftover from the prior test** (until the washing solutions of both chambers had a conductivity close to DI water, <1 µS/cm). Note that we repeated some tests like single $K^+$ permeation tests and other $K^+/Pb^{2+}$ mixed solution permeation tests with the same ratio to confirm that the tested $MoS_2$-COOH membrane was intact during the series of tests. The membranes were always kept in the wet state throughout the series of tests.

## Calculation of the Permeabilities/Flux ratio

As our previous studies show, ion concentration changes in the permeate side in the early stage show a linear increase versus time (**Supplementary Figs. 4b**,**c** and **6b,c** and **7b-d** and **9** and **26b,c** and **27 b,c** and **28a,b**). We can get the permeabilities of the ions, $P_i$ (mol/s), by calculating the slopes of linear profiles to make quantitative analysis, as well-established.

Given that every series tests of binary ions were done with one membrane, the relative flux ratio is the ratio of the permeabilities.

## Electrical Test

The setup for the electrical test is shown in **Supplementary Fig. 24**. Briefly, the $MoS_2$-COOH/PTFE complex membrane was mounted between two O rings (orifice ~ 1 mm) separating the two chambers. $PbCl_2$ solution (10 mM, ~ 200 μl) was first added into both chambers to convert the $MoS_2$-COOH membrane into $Pb^{2+}$-treated $MoS_2$-COOH membrane.

The I-V curves were recorded on a Bio-Logic VMP3 workstation using standard Ag/AgCl electrodes. The calculated conductance is the slope from the linear fitting of the measured I-V curves. CV tests with voltages ranging from 0 V to 0.1 V with a scan rate of 5 mV/s. Note that before every test, both chambers were thoroughly washed with DI water to remove salt leftover from prior test or treatment.

### The membrane uptake test

The $MoS_2$-COOH membranes (typical mass loading 0.2-0.3 mg) were put into 2.5 ml solution of either single salt solutions ($K^+$, $Co^{2+}$, $Ba^{2+}$, $Pb^{2+}$) or binary mixture solutions overnight. Then the membranes were taken out and thoroughly washed with flowing DI water (typically for ~ 2 min) to remove the ions solution remaining on the surface. Then the washed membranes were dissolved in 3ml $HNO_3$ (3%) under sonication overnight. Finally, the leaching $HNO_3$ solution was filtered and diluted for ICP-MS measurement.

### XRD measurement

The diffraction patterns in Bragg-Brentano geometry were obtained using a Rigaku benchtop Xray diffractometer equipped with HyPix-400 MF 2D hybrid pixel array detector (HPAD) and a Cu Kα X-ray source (1.5406 Å) operating at 40 kV and 15 mA. The interlayer spacing of the 2D channel is calculated according to the (002) peak position in the XRD spectra of $MoS_2$-COOH membranes.

### Scanning transmission electron microscopy (STEM) measurement

For the cross-section view imaging, the $MoS_2$-COOH membrane was embedded into Poly/Bed 812 resin and cut into ~ 90 nm thick slides using a ultramicrotome (Ultracut E, Reichert-Jung). Scanning transmission electron microscopy experiments were carried out using aberration corrected JEOL ARM 200CF with cold field emission gun (200kV).

STEM imaging and EDS mapping in **Supplementary Fig. 2** were conducted on aberration corrected Themis Z (S)TEM (Thermo Fisher Scientific) operated at 80 kV. Energy-dispersive X-ray spectroscopy (EDS) mapping was performed using a continuous scan with a dwell time of 2 μs and a beam current of 20 pA. The total acquisition time was about 30 min.

### All-atom MD simulations

**Computational Method:** We performed all-atom Molecular dynamics (MD) simulations to study the cooperative transport of mixed ($Pb^{2+}/K^+$) ions in the $MoS_2$-COOH membranes. Our model (**Fig. 3a**) consists of a bilayer $MoS_2$ sheet acting as a piston wall to exert external pressure, an ion-filled feed region, a multi-layered $MoS_2$ membrane serving as an ion channel, and a pure water permeate region [40]. The multi-layered $MoS_2$ membrane can be visualized as extracting two layers of $MoS_2$ from bulk crystals (2H-$MoS_2$) and being modified with corresponding acetate functional groups, while leaving a 11.4 Å interlayer spacing (Mo-Mo distance) that is consistent with the experiment ($Pb^{2+}$ locked membrane). We built a rectangular box with dimensions 170 Å × 57.2 Å × 54.0 Å, we chose the OPC3 water model, and visualized the model using Visual Molecular Dynamics (VMD)[41]. We used the parameters taken from Heinz[42] for the $MoS_2$ membrane. We applied the general AMBER force field (GAFF)[43] for the -COOH functional groups. We selected the 12-6-4

LJ-type nonbonded model[44] for the monovalent/divalent ions with the following potential function:

$$U_{MW}(r_{MW}) = \frac{C_{12}^{MW}}{r_{MW}^{12}} - \frac{C_6^{MW}}{r_{MW}^6} - \frac{C_4^{MW}}{r_{MW}^4} + \frac{e^2 Q_M Q_W}{r_{MW}} = \varepsilon_{MW}\left[\left(\frac{R_{min,MW}}{r_{MW}}\right)^{12} - 2\left(\frac{R_{min,MW}}{r_{MW}}\right)^6\right] - \frac{C_4^{MW}}{r_{MW}^4} + \frac{e^2 Q_M Q_W}{r_{MW}}$$

(Equation 1)

Herein, M and W represent the metal ion and an atom inside a water molecule, respectively. $r_{MW}$ is the distance between M and W. e is the charge of a proton. $Q_M$ and $Q_W$ are charges of M and W, respectively. $R_{min,MW}$ is the distance between M and W when their LJ potential reaches its minimum, and $\varepsilon_{MW}$ is the corresponding well depth at this minimum. The C4 term is added to account for the ion-induced dipole interactions between the metal ion and water molecule. These parameters were optimized to reproduce the experimental hydration free energy (HFE), ion-oxygen distance (IOD), and coordination number (CN) values simultaneously. For more detailed optimization procedure, please refer to our previous work[45]. The parameters of the ions we used in this work are shown in **Supplementary Table 3**, the target/calculate values of the HFE, IOD, and CN in the first solvation shell for $Pb^{2+}$ are listed in **Supplementary Table 4**. Note that due to the lack of parameters in the 12-6-4 model of nitrate anion, we used $Cl^-$ in our simulation instead of the nitrate anion used in the experiment. We chose $Cl^-$ for two reasons: first, its diffusion coefficient is similar to that of nitrate anion [46]; second, it is a commonly used anion in MD simulations[11]. In addition, calculation of the radial distribution function (RDF) of $Pb^{2+}$ and $Cl^-$ in the bulk solution shows that PbCl does not form ion clusters in the simulation (see **Supplementary Fig. 11**).

Molecular dynamics (MD) simulations were performed using the OpenMM package[47] and the settings of the **$K^+$/$Pb^{2+}$** permeation test were consistent with the experimental settings (**Supplementary Table 5**). We kept the $MoS_2$-COOH functional group in the Test 1 fully protonated but starting from Test 2, we deprotonated half of the $MoS_2$-COOH groups, as suggested by the infrared (IR) vibrational spectroscopy results that $Pb^{2+}$ can deprotonate the -COOH function group (**Ref. 35** of main text). Experiments have shown that the $MoS_2$-COOH membrane soaked with $Pb^{2+}$ will be locked by $Pb^{2+}$(fixed interlayer distance), so we put 18 $Pb^{2+}$ in the membrane of the Test 3-Test 9 system to neutralize and lock the negatively charged membranes, where the initial position of $Pb^{2+}$ in the membrane is taken from the equilibrium state of Test 2 after applying a force of 400 kcal/mol/nm (see snapshots in **Supplementary Fig. 12**). We also conducted similar permeation test for the $Co^{2+}$/$Pb^{2+}$ system (**Supplementary Table 8**).

All the systems were subject to a 10000-step energy minimization followed by a 5 ns NVT ensemble at 300 K using Langevin dynamics to equilibrate the system. The piston wall and membrane are positionally restrained, whereas ions are restrained to pass through the channel by applying a flat-well potential. We then removed the constraints on the ions and performed the non-equilibrium production simulations in the NVT ensemble where different external pressures (**Supplementary Table 5** and **Supplementary Table 8**) were applied on the piston wall to simulate ions with different ratios permeating through the $MoS_2$-COOH membrane[40,48]. The temperature was kept at 300 K using a Langevin thermostat with a friction coefficient of 1 $ps^{-1}$. The particle mesh Ewald method[49] was employed to calculate electrostatic interactions with a short-range cutoff of 1.0 nm. Periodic boundary conditions were employed, and with a simulation time step of 2 fs. To speed up the MD simulations and collect sufficient statistics in the simulations at the ns scale, we employed external pressures higher than the experimentally measured osmotic pressure (~0.5 MPa). Atomic coordinates are saved every 20 ps.

To validate the accuracy of the ion parameters we used in this work, we applied steered molecular dynamics (SMD) simulations to calculate the potential of mean force (PMF) of a single $Pb^{2+}$ or $Co^{2+}$ or $K^+$ ion passing through the $MoS_2$-COOH membrane. We first construct a 4 layer $MoS_2$-COOH membrane with dimensions 55.0 Å × 57.2 Å × 30.0 Å, while leaving interlayer spacings that are consistent with experiment. We then placed the membrane in a rectangular box of size 100.0 Å × 57.2 Å × 30.0 Å and solvated with 2904 OPC3 water molecules. We kept the $MoS_2$-COOH functional group in the $K^+$ system fully protonated while half deprotonated in the $Pb^{2+}$ and the $Co^{2+}$ system. 18 $Pb^{2+}$ and 18 $Co^{2+}$ ions were added to neutralize the $Pb^{2+}$ system and the $Co^{2+}$ system, respectively. For each simulation, we added 1 cation and a corresponding number of $Cl^-$ ions to neutralize the system, where we placed the cation 10 Å outside the membrane (**Supplementary Fig. 13a**). Following the equilibration procedure mentioned above, we performed an SMD simulation for 40 ns where the cation position was increased gradually from an initial -10 Å to a final 30 Å. We used harmonic restraints with a force constant 50 kcal/mol to gradually move the cation from outside of the membrane (distance = –10 Å) to the inside of the membrane. We then used the steered MD trajectories as initial coordinates for our umbrella sampling windows. We used the distance as our reaction coordinate, and we divided the trajectories into 40 umbrella windows with 0.1 nm intervals (-10 Å – 30 Å). We then applied a harmonic potential with force constant 20 kcal/mol to constrain each window to the specific distance along the reaction coordinate. Each umbrella window was simulated for 10 ns, where the first 4 ns was discarded as equilibration. We then calculated the potential of mean force using the weighted histogram analysis method (WHAM)[50], as shown in **Supplementary Fig. 13b**.

Here the diffusion coefficient D we calculated is obtained by fitting the MSD with respect to the lag-time to a linear model: $D_d = \frac{1}{2d} \lim_{t \to \infty} \frac{d}{dt} MSD(r_d)$, where d is the dimensionality of the MSD (d=3). MSD was computed from the following expression, known as the Einstein formula:

$$MSD\ (r_d) = \langle \frac{1}{N} \sum_{i=1}^{N} |r_d - r_d(t_0)|^2 \rangle_{t_0} \qquad \text{(Equation 2)}$$

where N is the number of equivalent particles the MSD is calculated over, r are their coordinates and d the desired dimensionality of the MSD.

**Data availability**
All data are in the main text or supplementary materials.

**Acknowledgments**
This work is supported by Advanced Materials for Energy-Water-Systems (AMEWS) Center, an Energy Frontier Research Center funded by the U.S. Department of Energy, Office of Science, Basic Energy Sciences. Cross-sectional STEM experiments used instruments in the Electron Microscopy Core of UIC's Research Resources Center. Other STEM experiments were carried out in part in the Materials Research Laboratory Central Research Facilities, University of Illinois.

## Author contributions

M.W. and C.L. conceived the project. M.W. designed the experiments, performed the bulk of the permeation/electrical/uptake/XRD tests and analyzed the data. Q.X. designed the MD model, performed the MD simulations and analyzed the data. G.Y. contributed to ICPMS tests. Y.H. performed cross-sectional STEM imaging. X.Y. conducted SEM and AFM analysis. Z.Lyu and Q.C. performed STEM imaging and EDS mapping. E.H. and K.X. contributed to electrical tests. L.S. contributed to ICPMS tests. Z.Li and K.M.M. Jr contributed to the parameterization of the MD model. N.H.C.L contributed to IR measurements and analysis. M.W., Q. X., G.C.S. and C.L. wrote the manuscript with inputs from all the authors. G.C.S. and C.L. supervised the project.

## Competing interests

The authors declare no competing interests.

## Additional information

**Supplementary information** The online version contains supplementary material available at..
**Correspondence and requests for materials** should be addressed to George C. Schatz or Chong Liu.

Supplementary Information for

**Cooperative and Inhibitory Ion Transport in Functionalized Angstrom-scale Two-dimensional Channels**

by Mingzhan Wang et al.

**This PDF file includes:**

Supplementary Figures 1-31

Supplementary Tables 1-8

Supplementary References 1-6

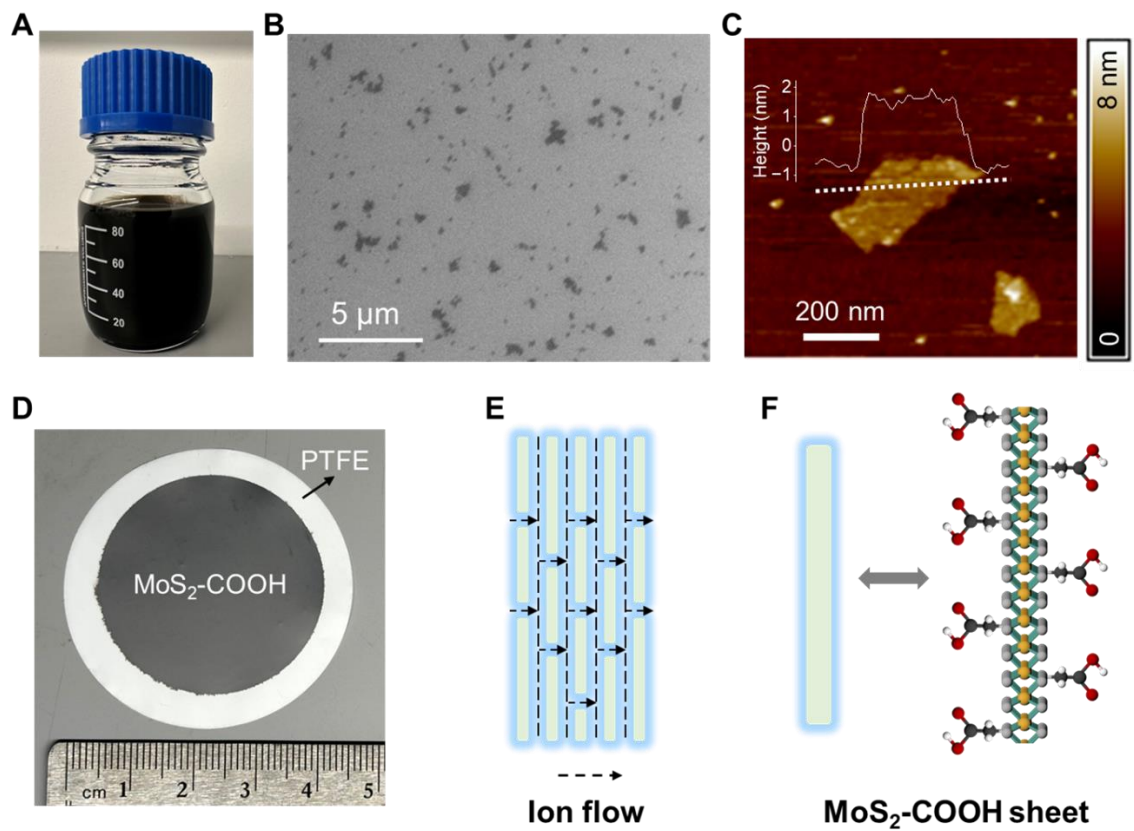

**Supplementary Figure 1: MoS$_2$-COOH nanosheet solution and membrane.** (**A**) MoS$_2$-COOH nanosheet dispersion solution in water. (**B**) Typical SEM image of MoS$_2$-COOH nanosheet dispersed on Si substrate. (**C**) Typical AFM image of MoS$_2$-COOH nanosheet dispersed on Si substrate, which shows a bilayer MoS$_2$-COOH nanosheet. (**D**) Photo of the prepared MoS$_2$-COOH membrane on PTFE substrate (**Fig. S8** below). (**E**) Schematic illustration of the ion pathway in the 2D channels. (**F**) Illustration of the structure of the MoS$_2$-COOH nanosheet.

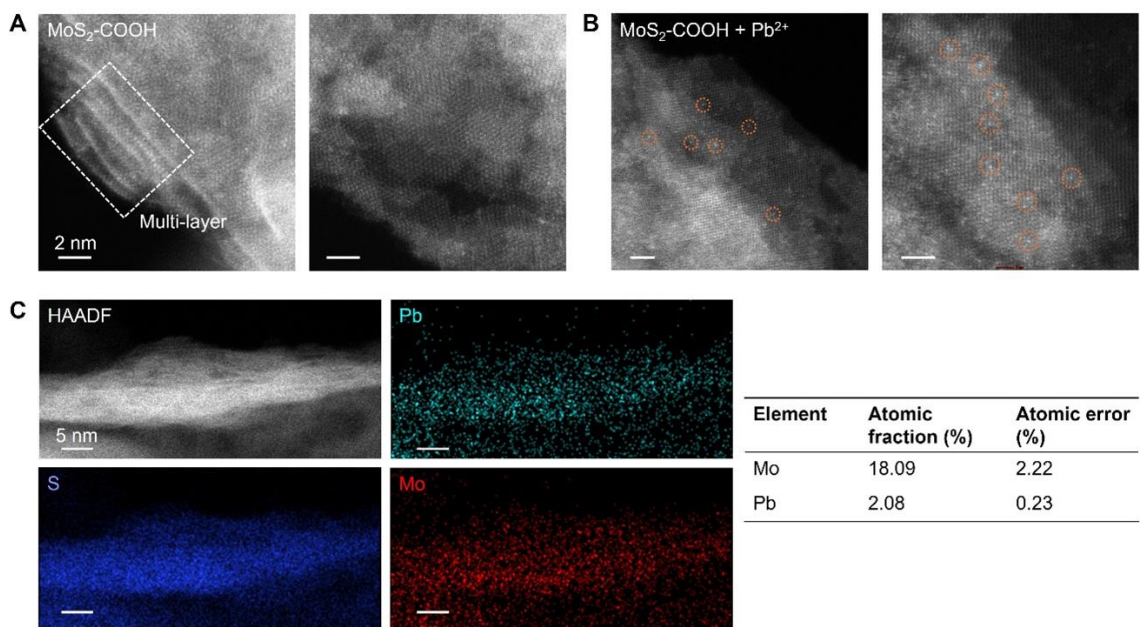

**Supplementary Figure 2: Electron microscopy characterization of the MoS$_2$-COOH membrane before and after treatment with Pb$^{2+}$.** (**A**, **B**) High-angle annular dark-field STEM (HAADF-STEM) images showing (**A**) the layered structure of MoS$_2$-COOH membrane before ion uptake and (**B**) the membrane after treatment with Pb$^{2+}$. Orange circles represent the adsorbed Pb$^{2+}$ cations. (**C**) Energy-dispersive X-ray spectroscopy (EDS) of the membrane confirming the presence of Pb$^{2+}$ after ion uptake. The scale bars in (**A**) and (**B**) are 2 nm and those in (**C**) are 5 nm.

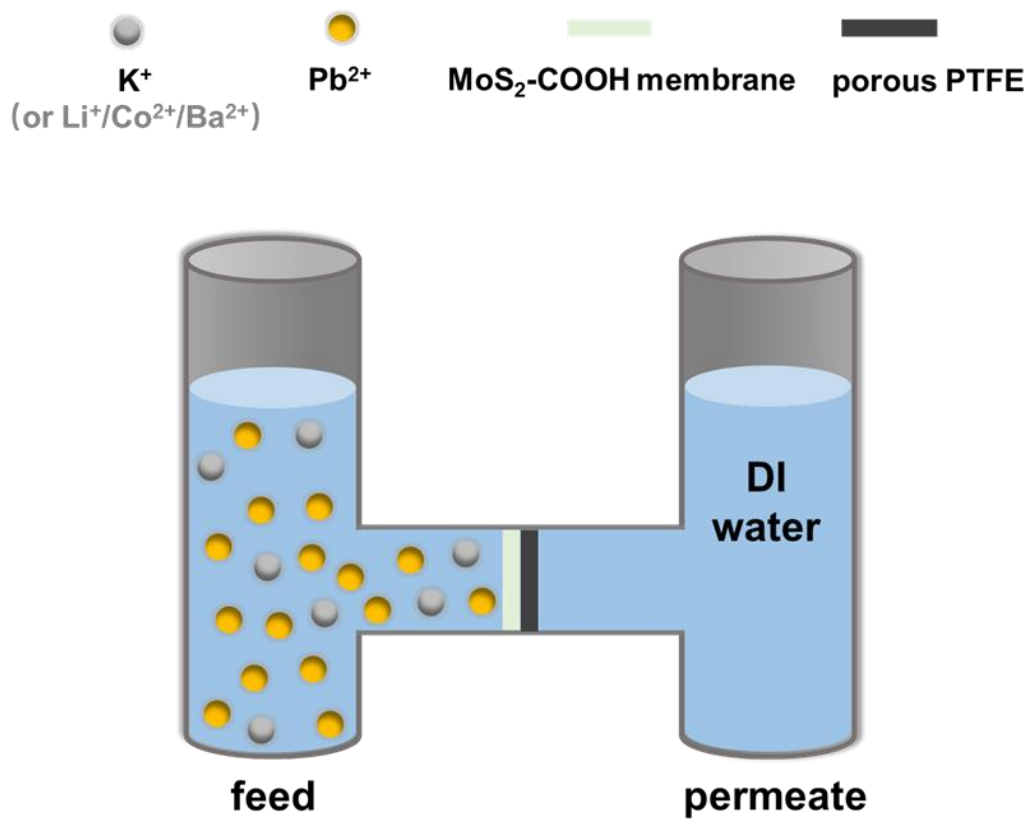

**Supplementary Figure 3: Schematic of the permeation test to measure the permeation rates of ions in the 2D MoS₂-COOH channels.** One chamber is the feed, and the other chamber is the permeate (i.e. deionized water of the same volume). The membranes to be tested were mounted in between the two chambers. Both chambers were strongly stirred to mitigate concentration polarizations during test.

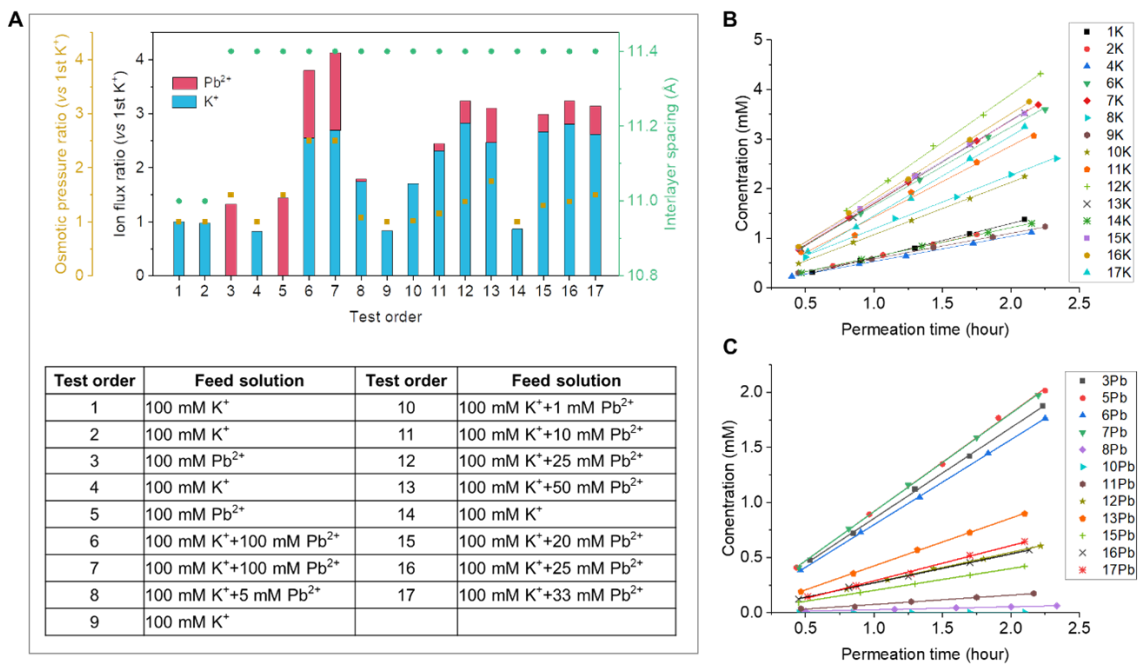

**Supplementary Figure 4: The sequential permeation tests of $K^+/Pb^{2+}$ (nitrate anion) across one $MoS_2$-COOH membrane.** (**A**) The order of the sequential permeation tests and the results. (**B**, **C**) show concentration profiles of $K^+$ and $Pb^{2+}$ in the permeate side, respectively. The detailed values are shown in **Table S2**. The number annotations in (**B**, **C**) correspond to test order in (**A**).

**Note:**

The highly repeatable results demonstrated in Test 4/9/14 and Test 12/16 not only demonstrate the integrity of membrane throughout the series of tests, but also unambiguously corroborate the enhanced transport of $K^+$ in the presence of $Pb^{2+}$ in the feed.

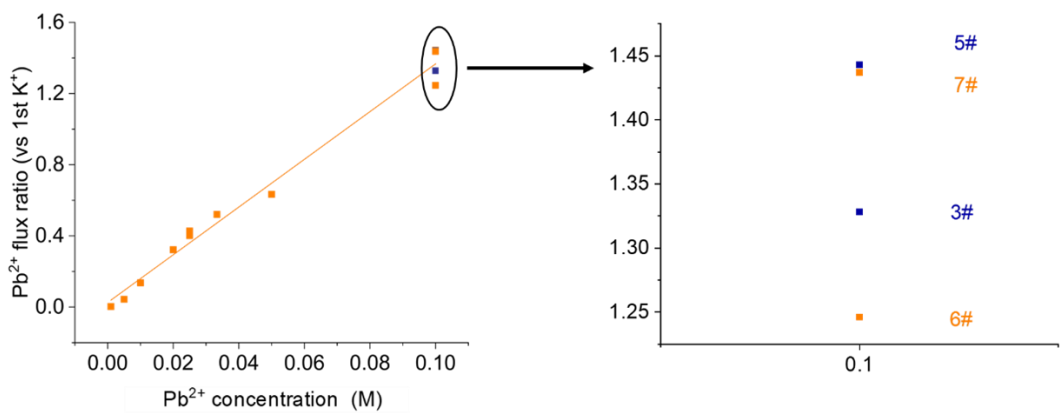

**Supplementary Figure 5: The permeation rate of $Pb^{2+}$ versus its concentration in the feed in the sequential permeation tests across the $MoS_2$-COOH membrane.** The number in the right figure indicates the test order. The blue squares (3# and 5#) correspond to tests with only $Pb^{2+}$ (100 mM) in the feed and orange squares (6# and 7#) correspond to tests with $K^+/Pb^{2+}$ (100 mM) in the feed. See **Table S2** for details. The linear fit in the left figure includes all data points in the plot.

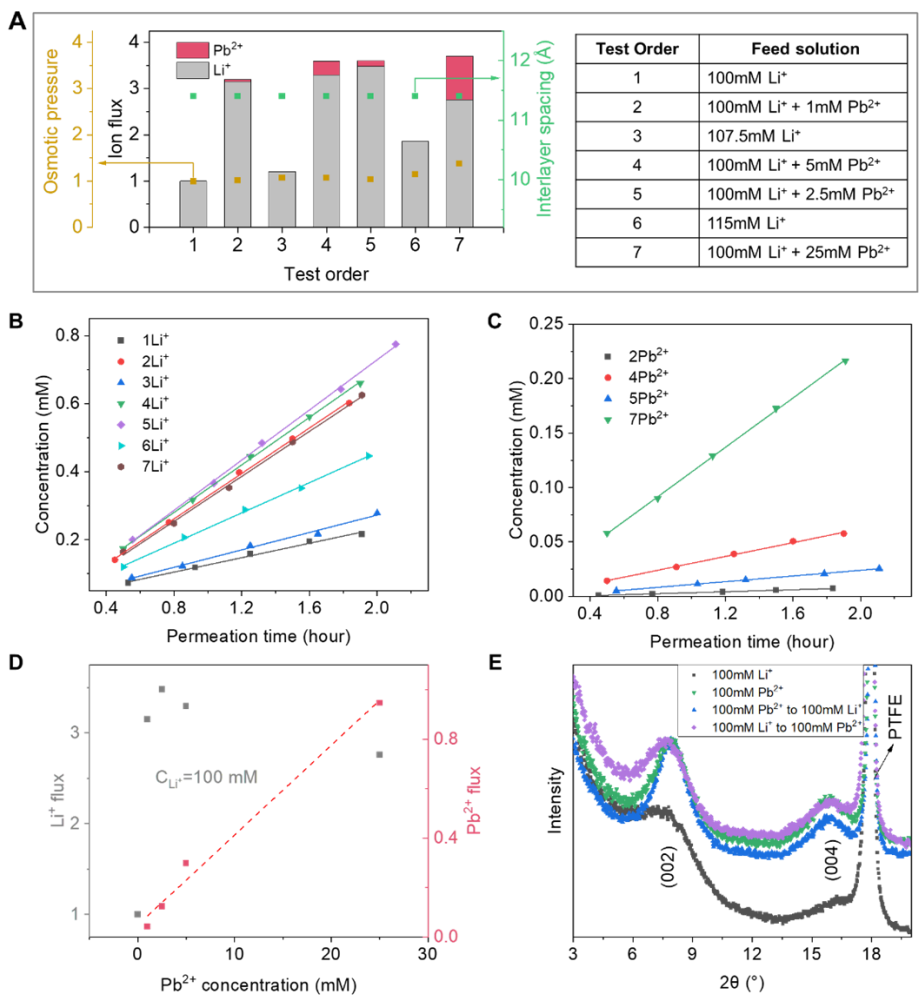

**Supplementary Figure 6: The sequential permeation tests of $Li^+/Pb^{2+}$ (nitrate anion) across one $Pb^{2+}$-locked $MoS_2$-COOH membrane.** (**A**) The order of the sequential permeation tests and the results. (**B**, **C**) show concentration profiles of $Li^+$ and $Pb^{2+}$ in the permeate side, respectively. The number annotations in (**B**, **C**) correspond to test order in (**A**). (**D**) Ion fluxes versus the concentration of $Pb^{2+}$ in the feed in the series of tests. All the fluxes are relative to that of $Li^+$ in the first test. (**E**) XRD spectra confirming the fixed interlayer spacing by $Pb^{2+}$ in the tests. The annotation "A to B" in (**E**) indicates that the $MoS_2$-COOH membrane was first saturated in solution A and then switched to solution B for saturate uptake, and finally the $MoS_2$-COOH membrane was taken out for XRD measurements.

**Notes:**

1. The $MoS_2$-COOH membrane was soaked in 100 mM $Pb(NO_3)_2$ to fix the interlayer spacing before starting the series of tests.

2. The results of Test 3 and 5 (Test 4 and 6), which have a same initial osmotic pressure (total molarity of ions), further confirms that the enhanced Li$^+$ flux is attributed to the presence of Pb$^{2+}$ in the feed solution.

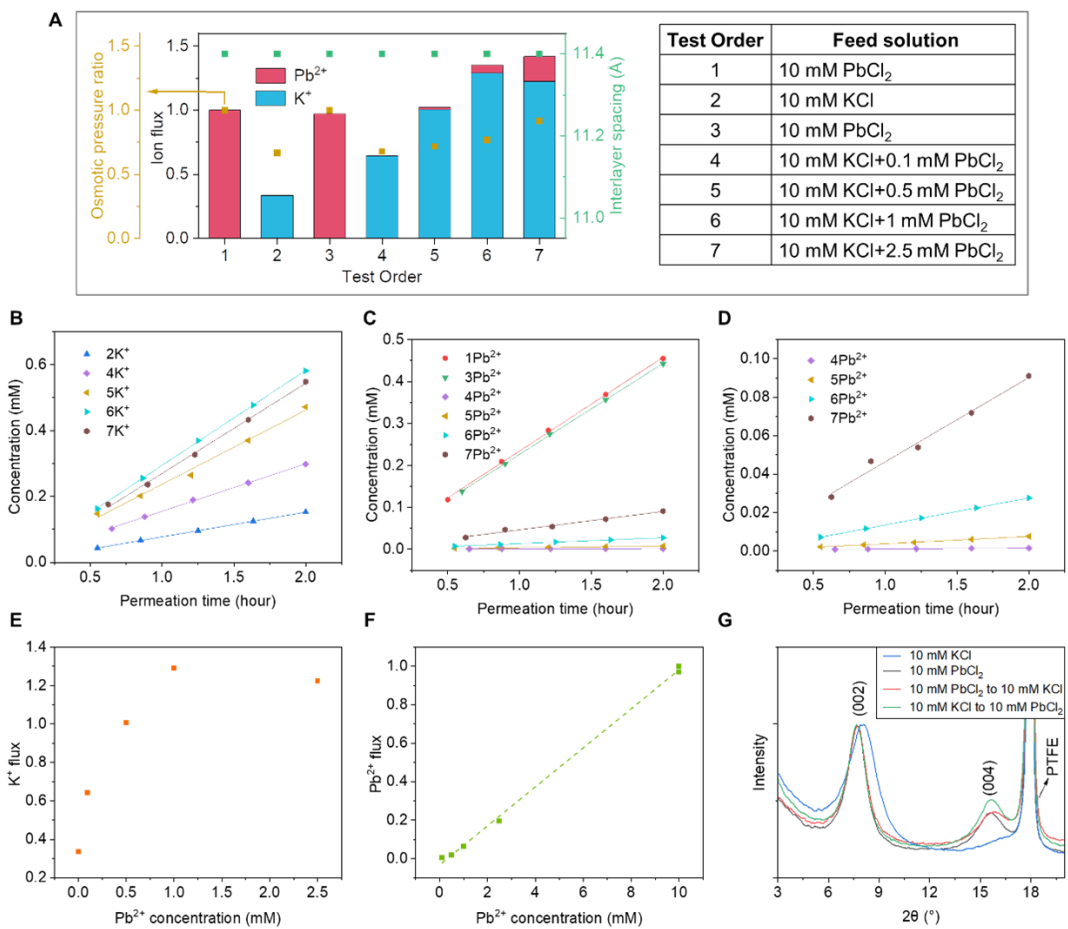

**Supplementary Figure 7: The sequential permeation tests of K$^+$/Pb$^{2+}$ (chloride anion) across one MoS$_2$-COOH membrane.** (**A**) The order of the sequential permeation tests and the results. (**B**, **C**, **D**) show concentration profiles of K$^+$ and Pb$^{2+}$ in the permeate side, respectively. (**D**) shows relevant zoomed-in plots in (**C**). The number annotations in (**B**-**D**) correspond to test order in (**A**). (**E**, **F**) The fluxes of K$^+$ and Pb$^{2+}$ versus the concentration of Pb$^{2+}$ in the feed in the series of tests, respectively. (**G**) XRD spectra confirming the fixed interlayer spacing by Pb$^{2+}$ in the tests. The annotation "A to B" in (**G**) indicates that the MoS$_2$-COOH membrane was first saturated in solution A and then switched to solution B for saturate uptake, and finally the MoS$_2$-COOH membrane was taken out for XRD measurements.

**Note:**

Given the low solubility of $PbCl_2$ in water (~36 mM), we set the concentration of $K^+$ in the series of tests using $Cl^-$ anion as ~ 10 mM to ensure that all species in the binary mixture solution are fully ionized.

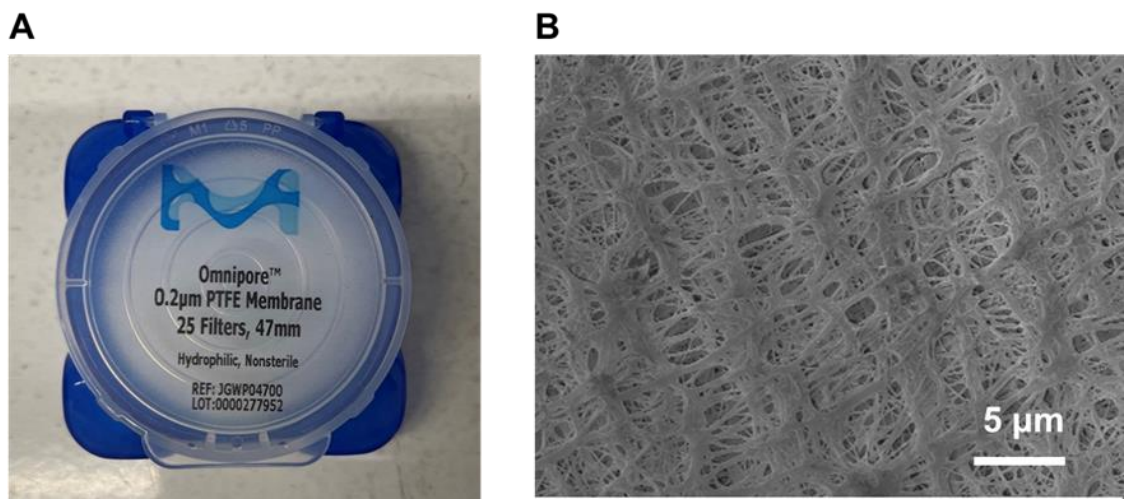

**Supplementary Figure 8: Ion transport across the PTFE membrane.** (**A**) Photo of the commercial PTFE membrane used. (**B**) SEM images of the PTFE membrane showing the large pores. 5 nm Pd/Pt were coated to eliminate the charging effect in SEM imaging.

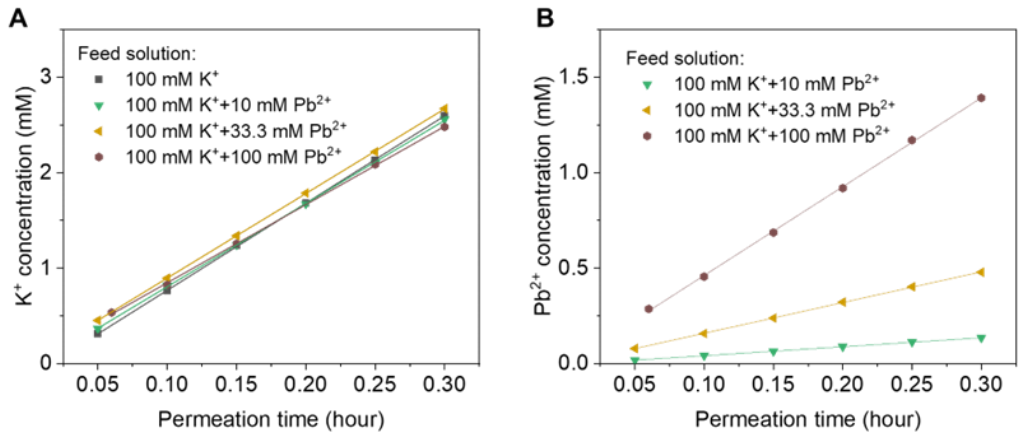

**Supplementary Figure 9: The permeation tests of $K^+/Pb^{2+}$ (nitrate anion) across one PTFE membrane.** (**A**, **B**) show representative concentration profiles of $K^+$ and $Pb^{2+}$ in the permeate side, respectively. The fluxes of $K^+$ in (**A**) almost keep the same, regardless of the presence of $Pb^{2+}$ in the feed or not.

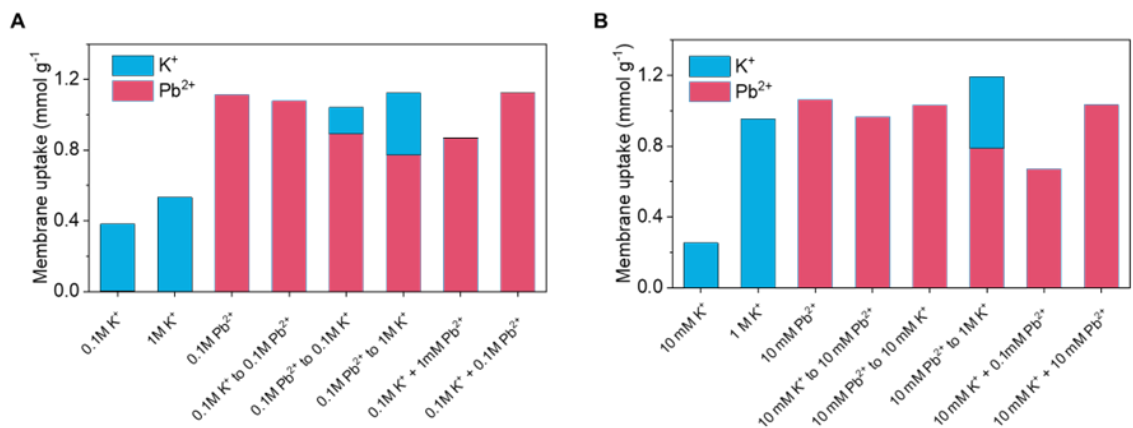

**Supplementary Figure 10: MoS$_2$-COOH membrane uptake results for K$^+$/Pb$^{2+}$ using nitrate anion (A) and chloride anion (B).** The annotation of "0.1M K$^+$ to 0.1M Pb$^{2+}$" in (**A**) means the MoS$_2$-COOH membrane was first soaked in 0.1M KNO$_3$ overnight for saturate K$^+$ uptake, and then the membranes were taken out of 0.1M KNO$_3$ and thoroughly washed with flowing DI water (typically for ~ 2 min) to remove the KNO$_3$ solution remaining on the membrane surface and finally the membrane was put into 0.1M Pb(NO$_3$)$_2$ overnight to measure the ion exchange from K$^+$ to Pb$^{2+}$ in the MoS$_2$-COOH membrane. Same interpretation can be made for similar annotations in the figures.

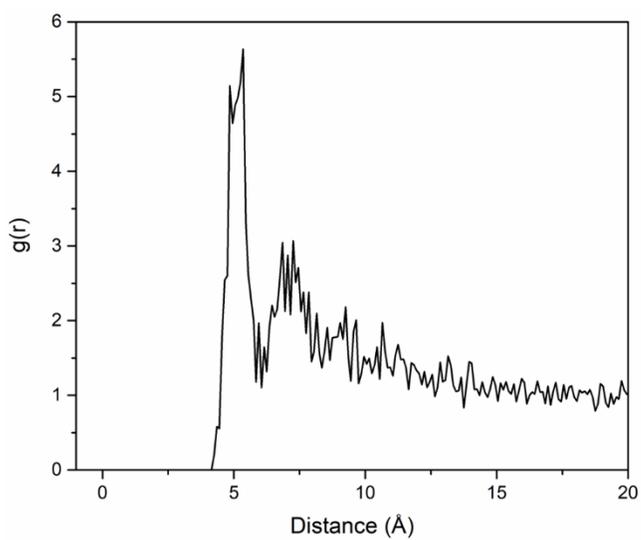

**Supplementary Figure 11: Radial distribution function (RDFs) between $Pb^{2+}$ and $Cl^-$ in bulk solution.**

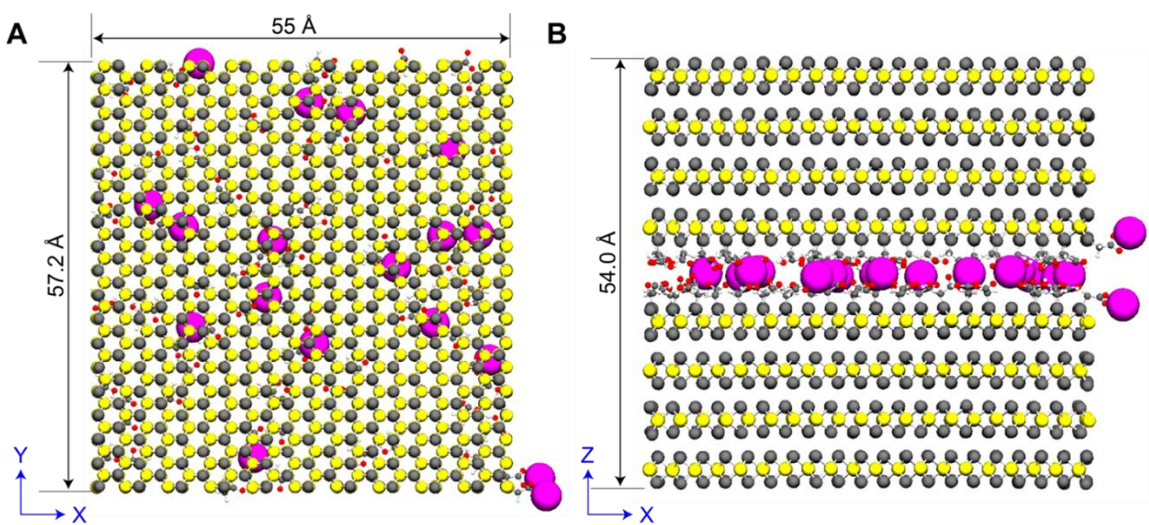

**Supplementary Figure 12: Snapshots of the initial position of Pb$^{2+}$ ions in the MoS$_2$-COOH membrane.** (**A**) and (**B**) are the top view and side view, respectively. The magenta beads in the membrane represent the Pb$^{2+}$ ions.

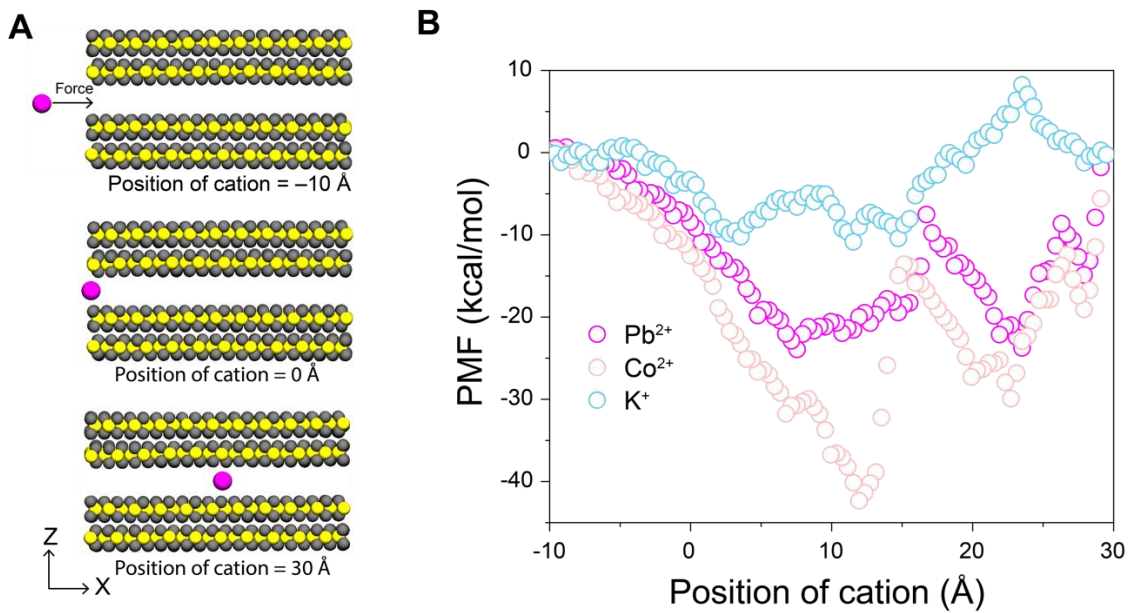

**Supplementary Figure 13: PMF calculations.** (**A**) Snapshots of a cation $Pb^{2+}/Co^{2+}/K^+$ that passing through $MoS_2$-COOH membranes. The -COOH group, other ions and water molecules are omitted for clarity. (**B**) PMF profiles of ions across the $MoS_2$-COOH membrane.

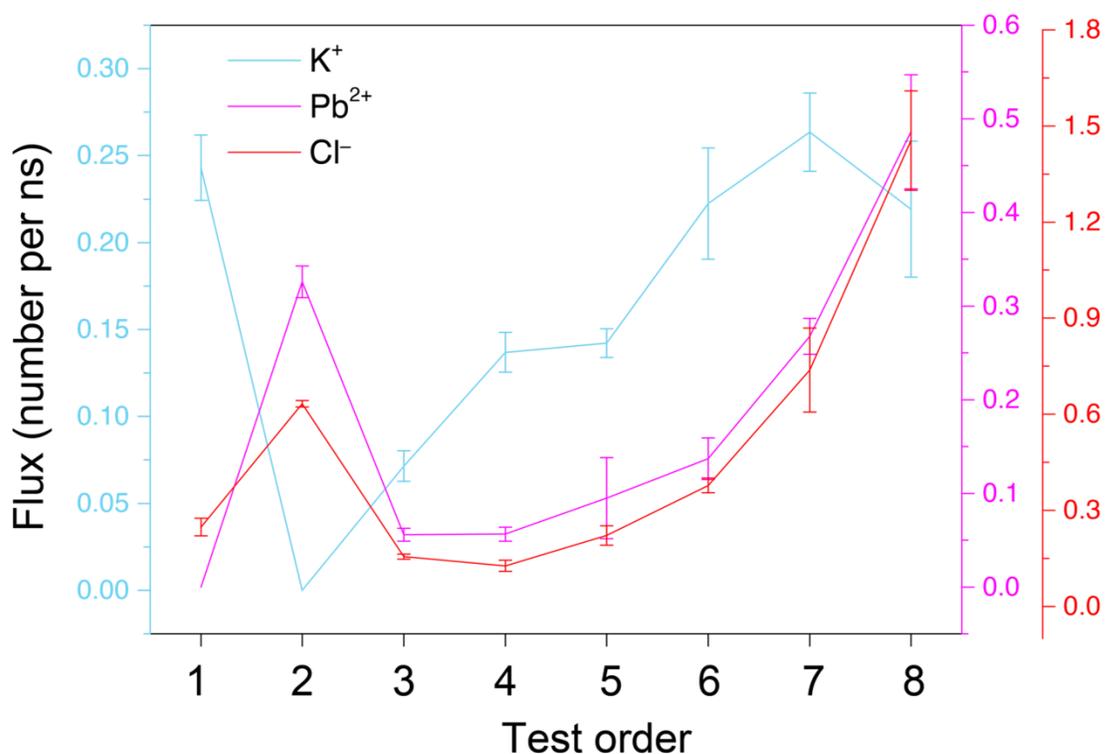

**Supplementary Figure 14: MD simulation results for the fluxes of $Pb^{2+}$, $K^+$ and $Cl^-$ across the $MoS_2$-COOH membrane under different order of the sequential permeation tests**. All the data are collected from the averaged values calculated from three independent simulations and the error bars are the standard error of means. We note that limited by the simulation scale, the observed $Pb^{2+}$ flux here includes $Pb^{2+}$ from the membrane, and therefore is much larger than what is observed in experiments. Nevertheless, our model provides a reliable qualitative explanation for the increase in $Pb^{2+}$ flux.

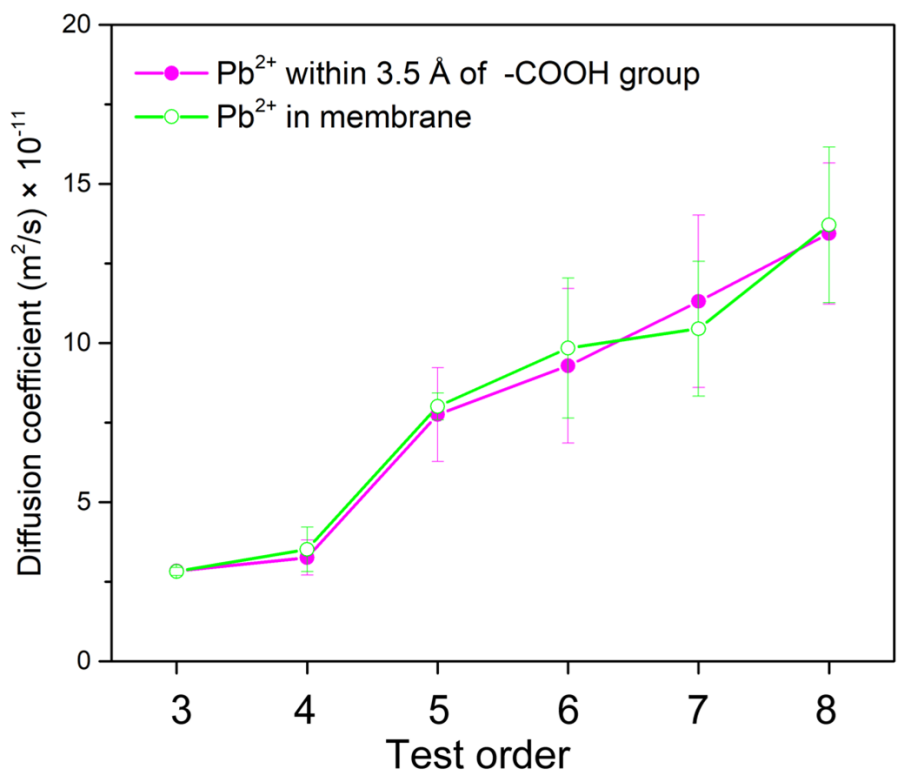

**Supplementary Figure 15: The diffusion coefficient of $Pb^{2+}$ within 3.5 Å of the -COOH group and in the membrane.** All the data are collected from the averaged values calculated from three independent simulations and the error bars are the standard error of means.

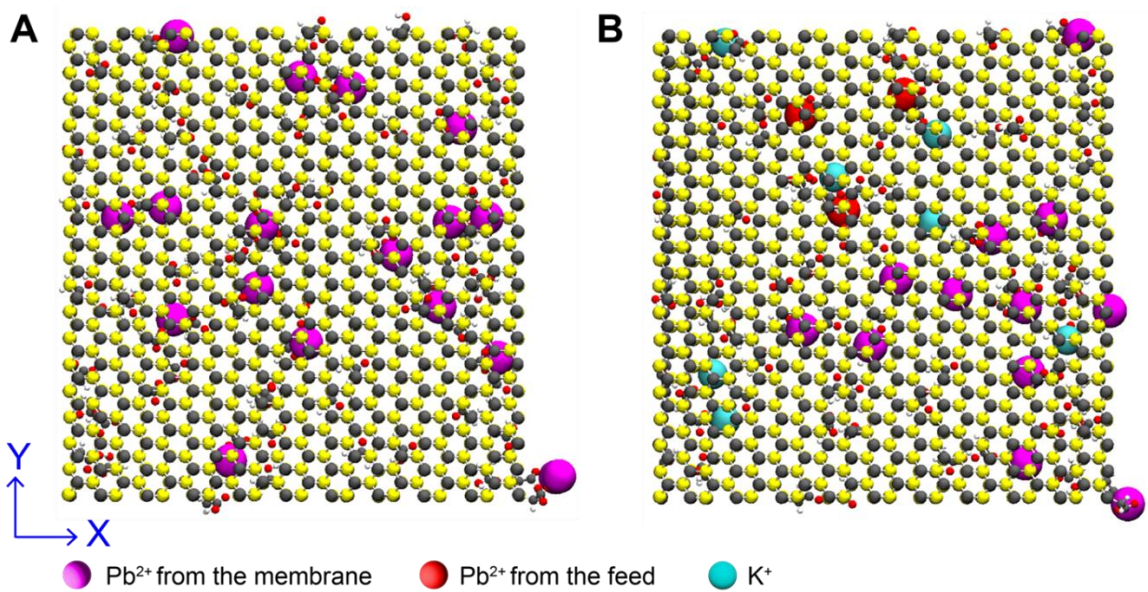

**Supplementary Figure 16: Snapshots of the positions of cations in the MoS$_2$-COOH membrane of Test 8**. (**A**) and (**B**) are the snapshots before and after applying pressure onto the piston wall, respectively.

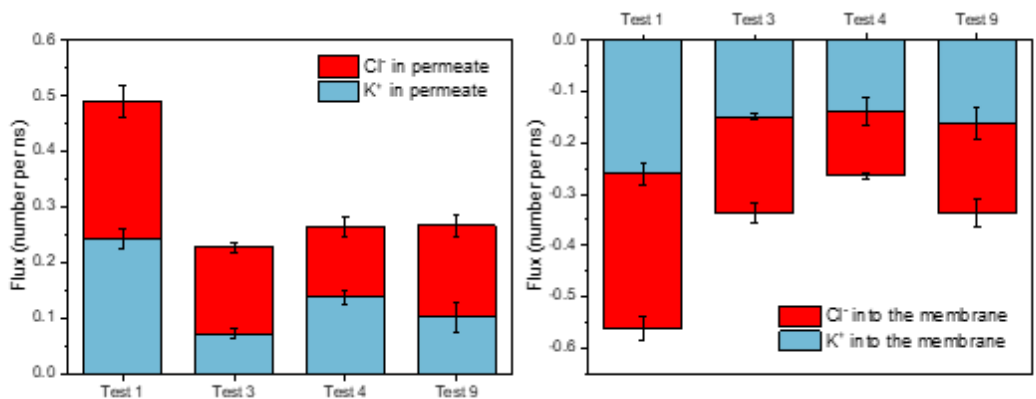

**Supplementary Figure 17: MD simulation results showing the fluxes of K$^+$ and Cl$^-$ across (A) and into (B) the MoS$_2$-COOH membrane.** All the data are collected from the averaged values calculated from three independent simulations and the error bars are the standard error of means.

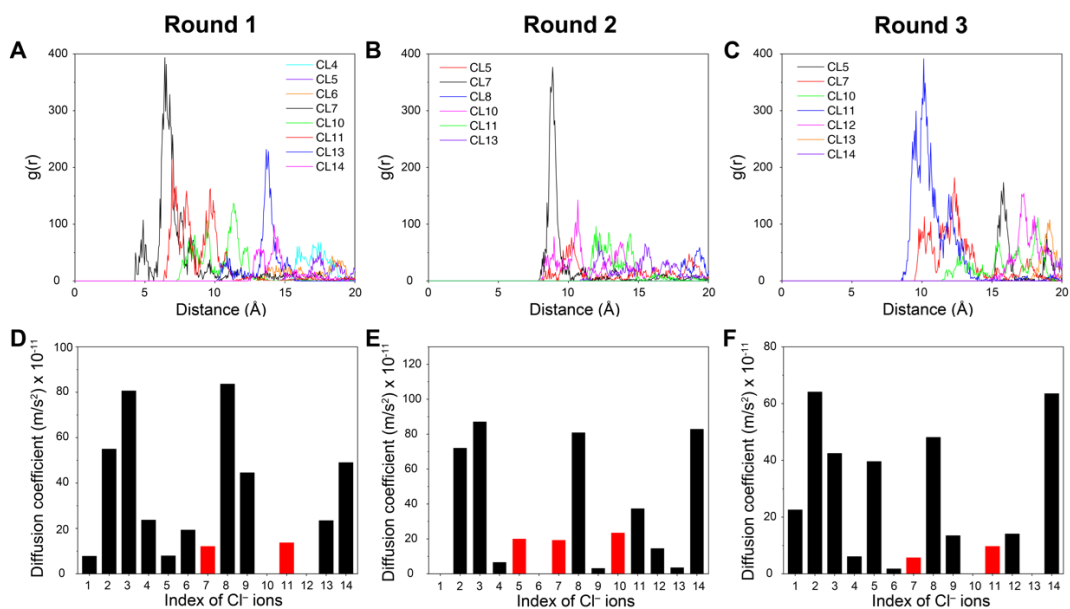

**Supplementary Figure 18: Radial distribution function (RDFs) calculations.** RDFs for $1^{st}$ (**A**), $2^{nd}$ (**B**) and $3^{rd}$ (**C**) rounds of the Test 4 system, where we calculate the RDFs between $Pb^{2+}$ and individual $Cl^-$ before entering the membrane of Test 4. We only show the RDFs for those $Cl^-$ within 20 Å of $Pb^{2+}$. The diffusion coefficient of each individual $Cl^-$ ion in the feed for $1^{st}$(**D**), $2^{nd}$ (**E**) and $3^{rd}$ (**F**) run of the Test 4 system. Here we only show the results of $Cl^-$ ions that can enter the membrane. For those $Cl^-$ never entering into the membrane, we consider their diffusion coefficient to be 0. The diffusion coefficient of $Cl^-$ ions that are within 10 Å of $Pb^{2+}$ are marked in red bars.

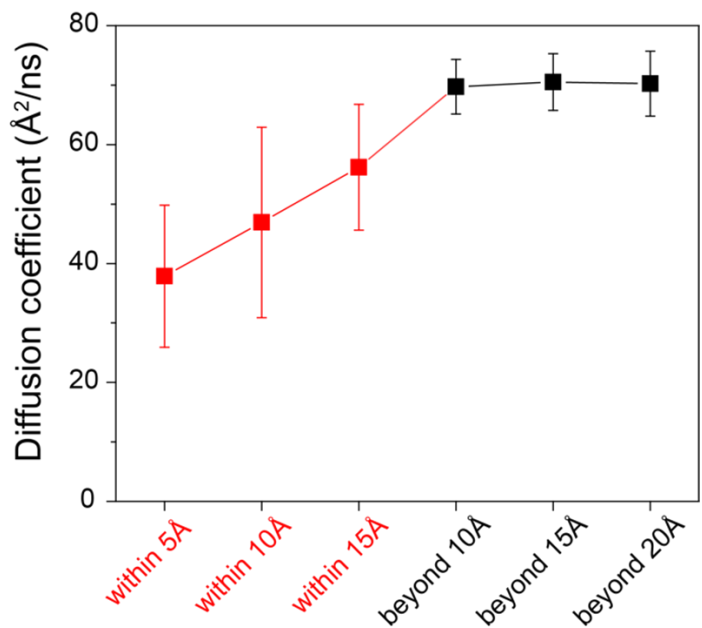

**Supplementary Figure 19: The diffusion coefficient of water that is within or beyond a certain range from $Pb^{2+}$ (while $Pb^{2+}$ is in the feed) for the Test 4 system.** All the data are collected from the averaged values calculated from three independent simulations and the error bars are the standard error of means.

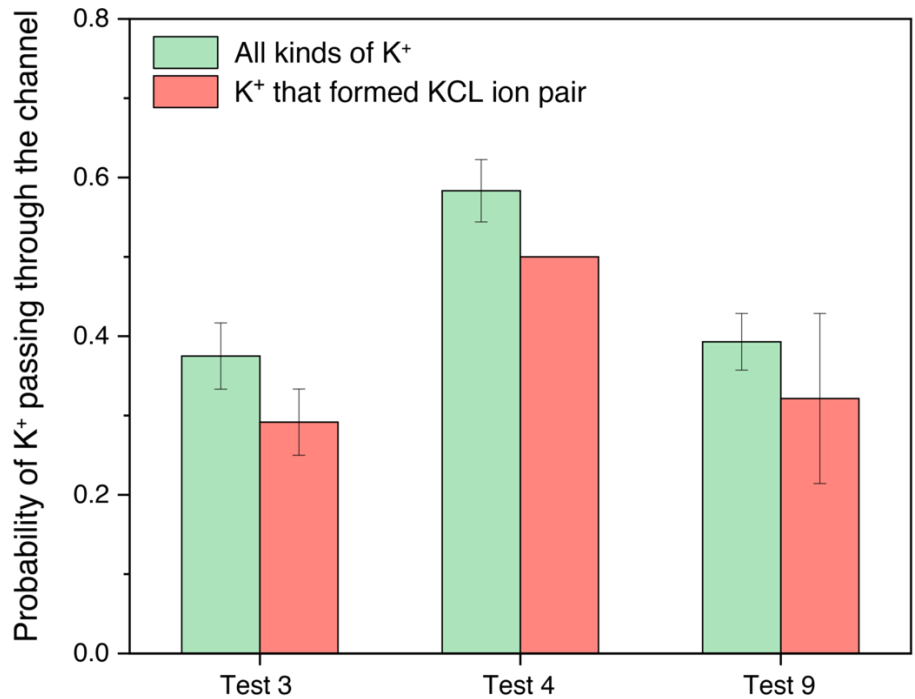

**Supplementary Figure 20: The probability of $K^+$ that pass through the membrane (pale greenish bar) and the ratio of $K^+$ that are knocked by $Cl^-$ and pass through the membrane (pink bar).** All the data are collected from the averaged values calculated from three independent simulations and the error bars are the standard error of means.

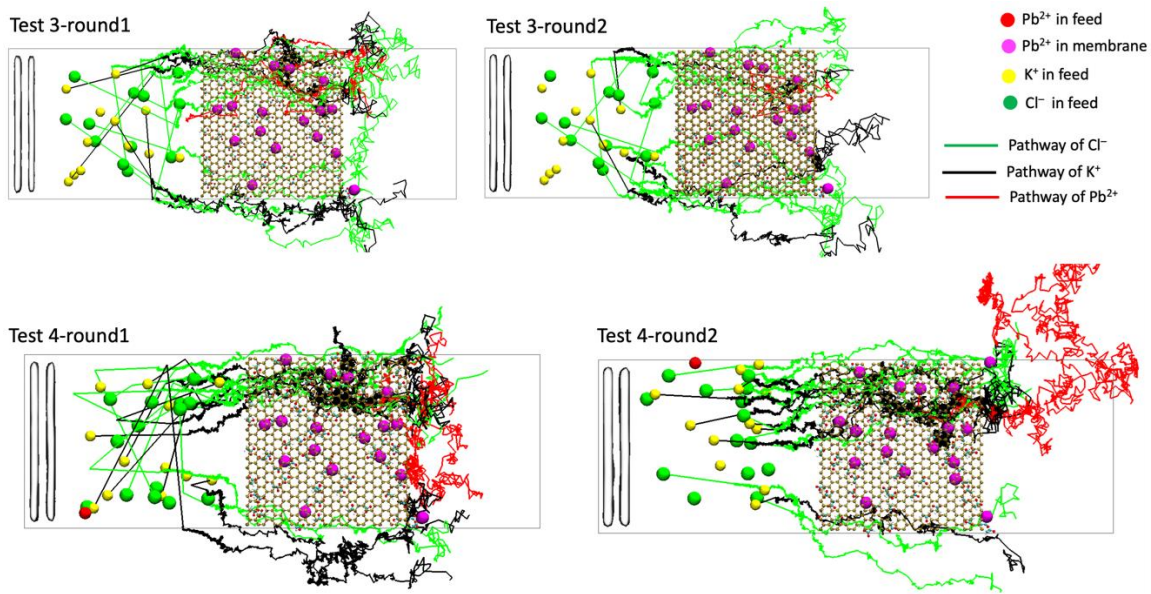

**Supplementary Figure 21: Pathways of ions into the permeate**. The green, black, and red lines represent the pathway of $Cl^-$ ions, $K^+$ ions and $Pb^{2+}$ ions, respectively. The simulation box is shown in a gray frame with Periodic boundary conditions (PBC).

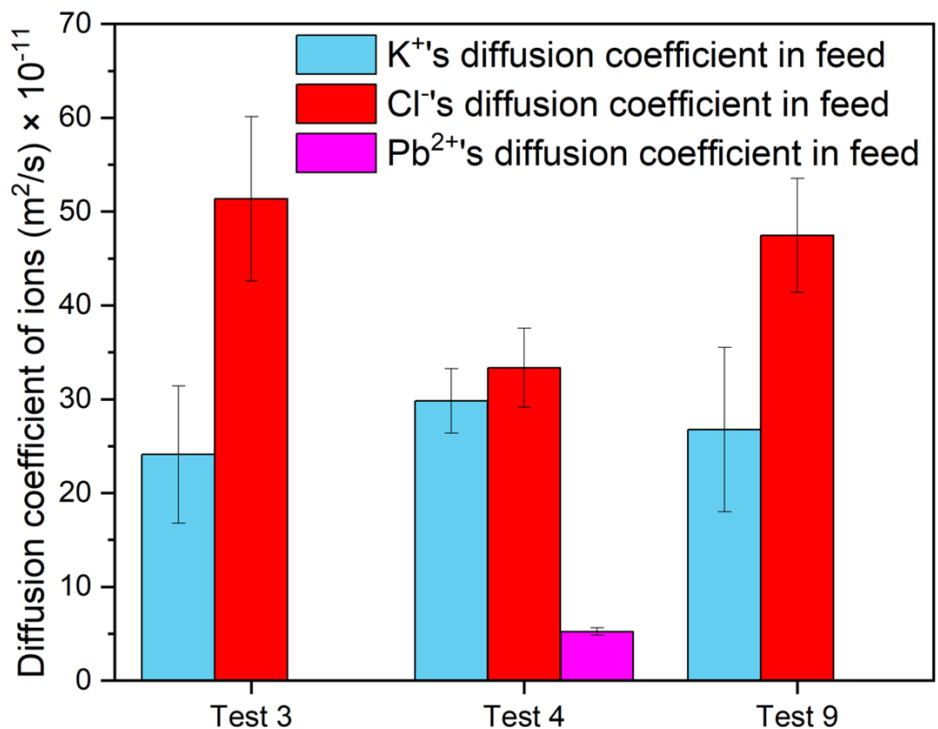

**Supplementary Figure 22: The diffusion coefficients of K$^+$, Cl$^-$, and Pb$^{2+}$ ions before entering the membrane.** Here we calculate the diffusion coefficients of each individual ion separately first, and then take average.

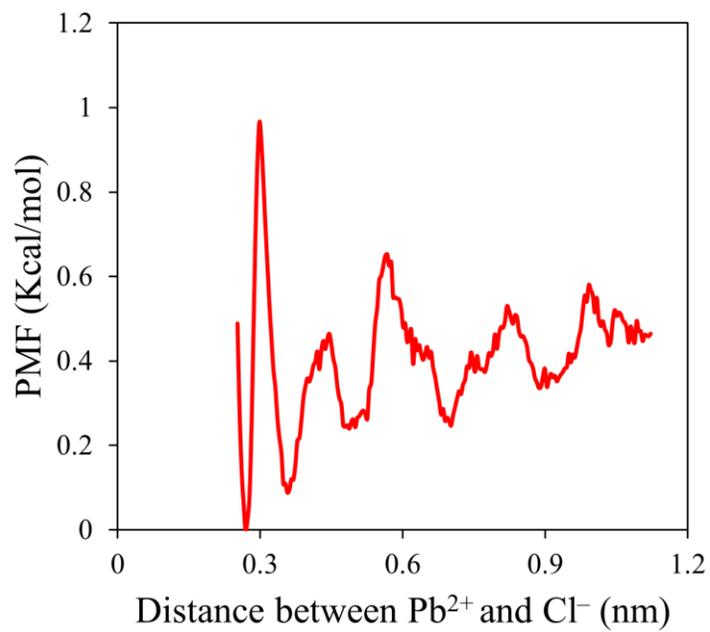

**Supplementary Figure 23: The potential mean force (PMF) between $Pb^{2+}$ and $Cl^-$ ions obtained from the bulk solution.**

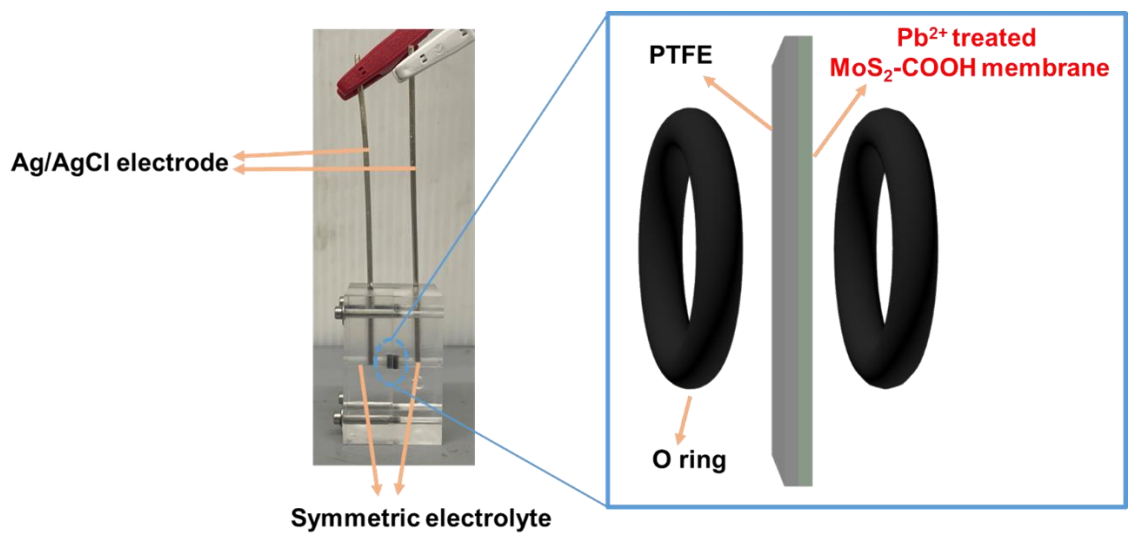

**Supplementary Figure 24: The setup for the electrical test.** Note: Before starting electrical tests, the MoS$_2$-COOH membrane was first soaked in 100 mM PbCl$_2$ to fix the interlayer spacing and was later thoroughly washed by DI water to remove PbCl$_2$ leftover, which is similar to the procedure of the sequential permeation tests.

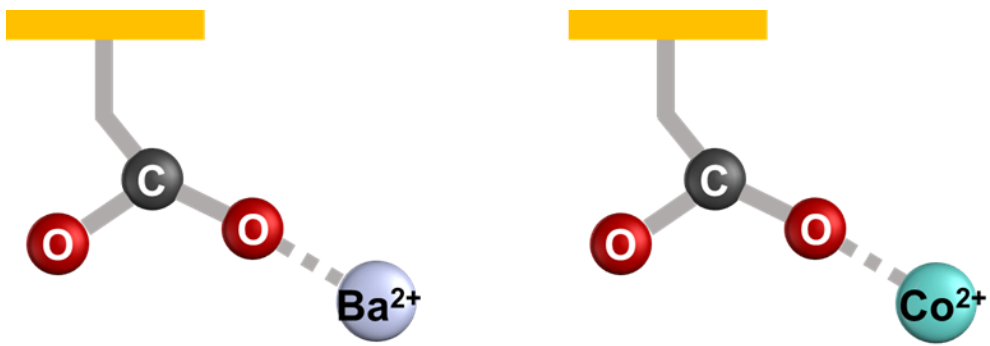

**Supplementary Figure 25:** Schematics showing the monodentate binding mode of $Co^{2+}$ and $Ba^{2+}$ with the acetate functional group in the 2D $MoS_2$-COOH channel.

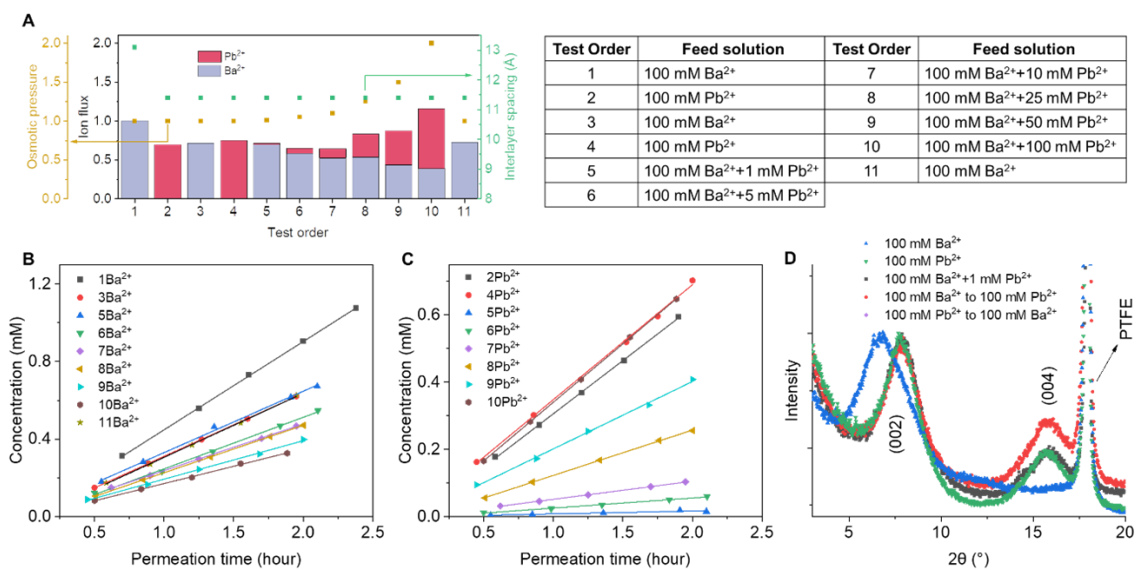

**Supplementary Figure 26: The sequential permeation tests of $Ba^{2+}/Pb^{2+}$ (nitrate anion) across one $MoS_2$-COOH membrane.** (**A**) The order of the sequential permeation tests and the results. The detailed values are shown in **Table S6**. (**B**, **C**) show concentration profiles of $Ba^{2+}$ and $Pb^{2+}$ in the permeate side, respectively. The number annotations in (**B**, **C**) correspond to test order in (**A**). (**D**) XRD spectra confirming the fixed interlayer spacing by $Pb^{2+}$ in the tests. The annotation "A to B" in (**D**) indicates that the $MoS_2$-COOH membrane was first saturated in solution A and then switched to solution B for saturated uptake, and finally the $MoS_2$-COOH membrane was taken out for XRD measurements.

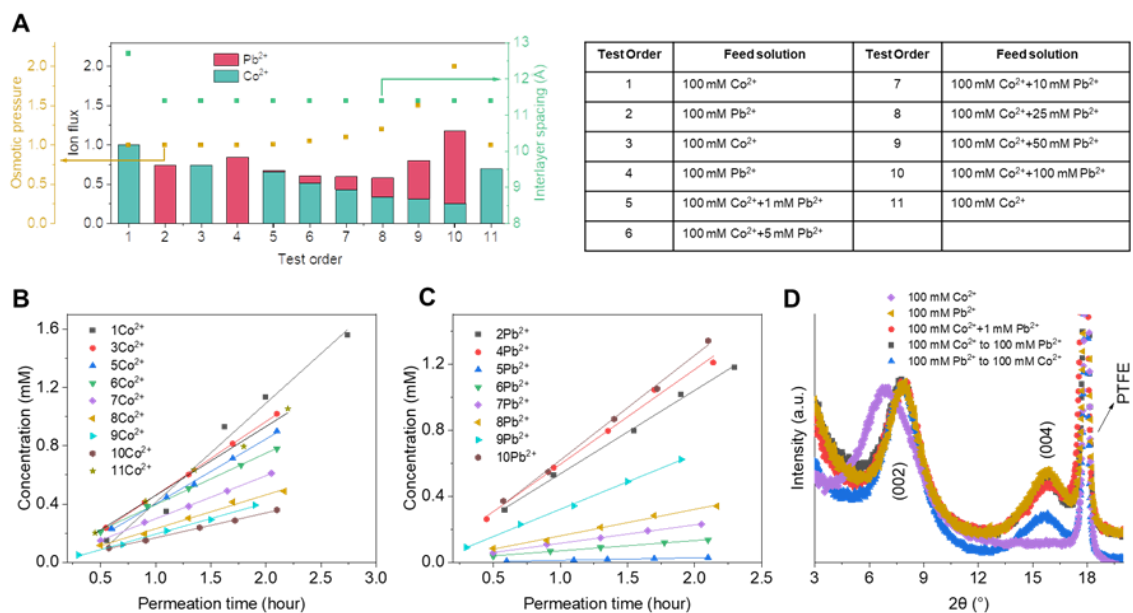

**Supplementary Figure 27: The sequential permeation tests of $Co^{2+}/Pb^{2+}$ (nitrate anion) across one MoS$_2$-COOH membrane.** (**A**) The order of the sequential permeation tests and the results. The detailed values are shown in **Table S7**. (**B**, **C**) show concentration profiles of $Co^{2+}$ and $Pb^{2+}$ in the permeate side, respectively. The number annotations in (**B**, **C**) correspond to test order in (**A**). (**D**) XRD spectra confirming the fixed interlayer spacing by $Pb^{2+}$ in the tests. The annotation "A to B" in (**D**) indicates that the MoS$_2$-COOH membrane was first saturated in solution A and then switched to solution B for saturated uptake, and finally the MoS$_2$-COOH membrane was taken out for XRD measurements.

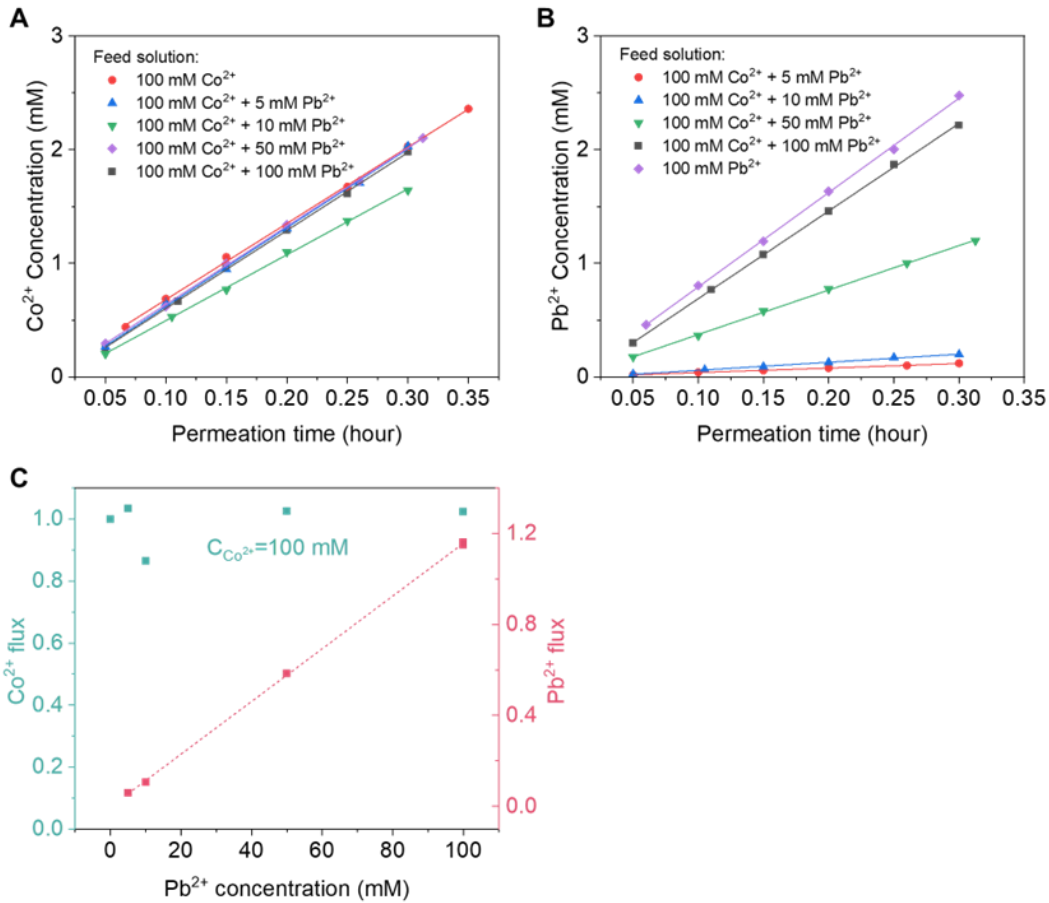

**Supplementary Figure 28: The sequential permeation tests of $Co^{2+}/Pb^{2+}$ (nitrate anion) across one PTFE membrane.** (**A**, **B**) show representative concentration profiles of $Co^{2+}$ and $Pb^{2+}$ in the permeate side, respectively. (**C**) Ion fluxes versus the concentration of $Pb^{2+}$ in the feed in the series of tests. The fluxes of $Co^{2+}$ throughout the series of tests almost keep the same, regardless of the presence of $Pb^{2+}$ in the feed or not.

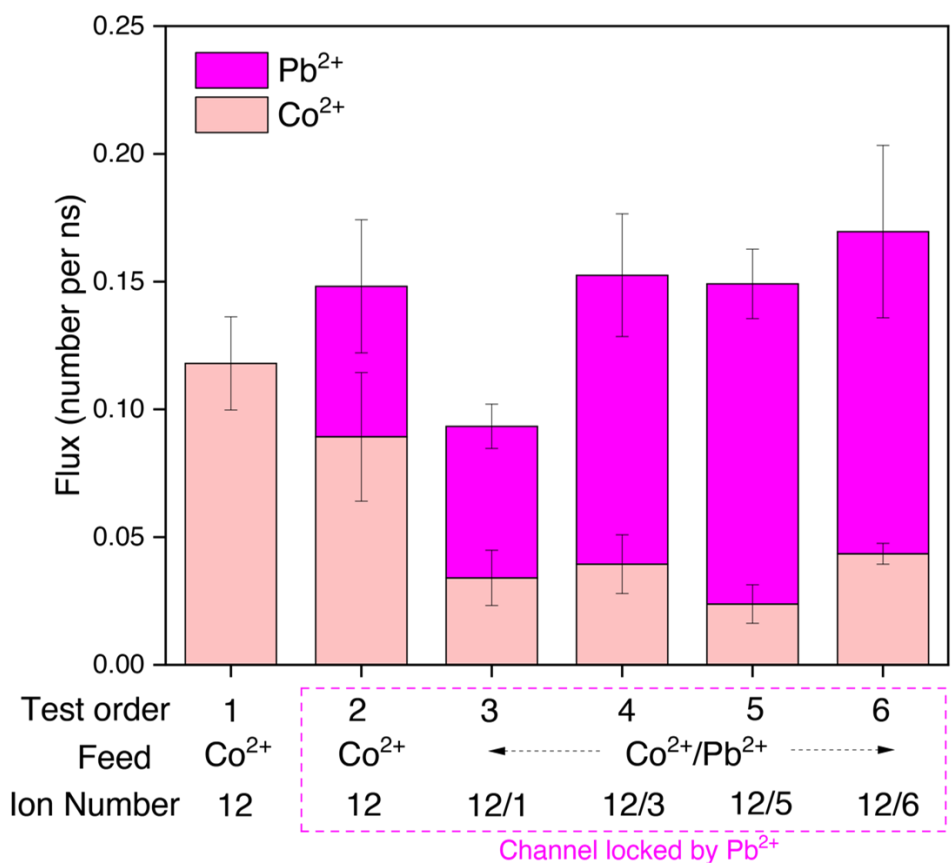

**Supplementary Figure 29: MD simulations results of the fluxes of $Pb^{2+}$ and $Co^{2+}$ across the $MoS_2$-COOH channel with different entries with $Co^{2+}/Pb^{2+}$ concentration ratios.**

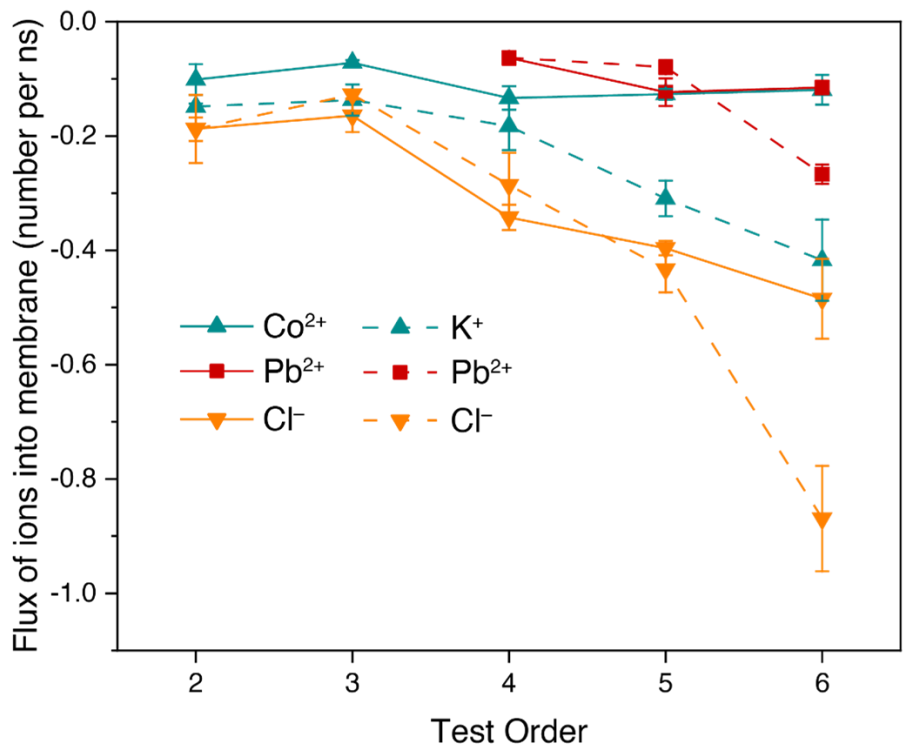

**Supplementary Figure 30: MD simulations results of the fluxes of ions entering the membrane.** Solid lines and dashed lines represent $Co^{2+}/Pb^{2+}$ and $K^+/Pb^{2+}$ tests, respectively.

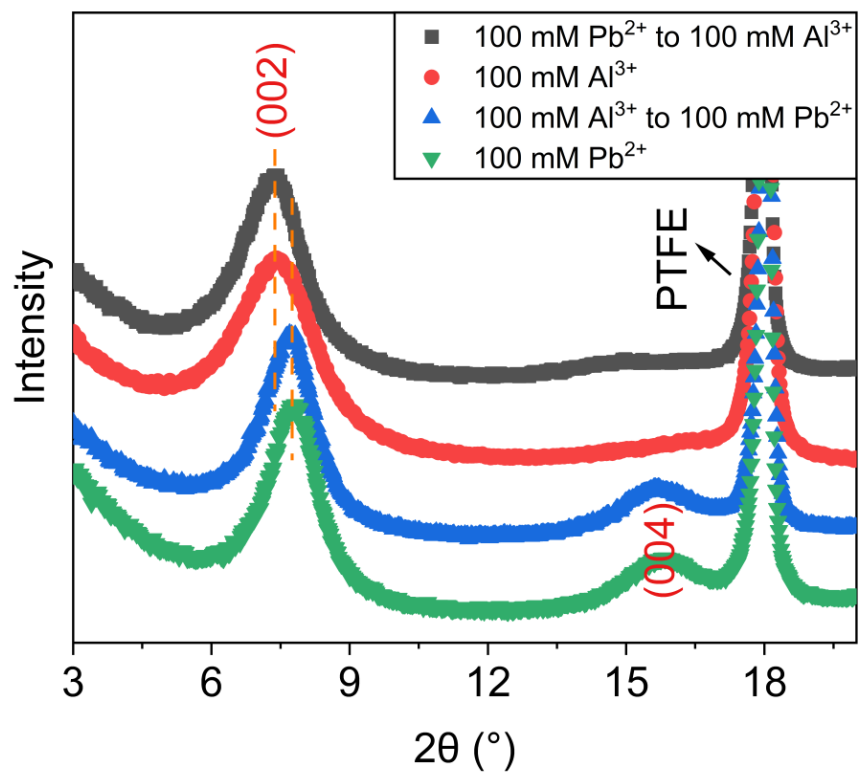

**Supplementary Figure 31: Dynamic change of the interlayer spacing in the MoS$_2$-COOH channels.** XRD spectra show that Pb$^{2+}$ and Al$^{3+}$ can replace each other in the MoS$_2$-COOH 2D channels, which is concomitant with a dynamic change in the interlayer spacing. Both salts are nitrate based.

**Supplementary Table 1: Relevant ion parameters in the aqueous bulk solution.**

|  | Hydration radius (Å) | Hydration enthalpy (- kJ mol$^{-1}$) | Bulk diffusion coefficient ($10^{-5}$ cm$^2$ s$^{-1}$) |
|---|---|---|---|
| **Li$^+$** | 3.82 | 531 | 1.029 |
| **K$^+$** | 3.31 | 334 | 1.957 |
| **Mg$^{2+}$** | 4.28 | 1949 | 0.706 |
| **Ba$^{2+}$** | 4.04 | 1332 | 0.847 |
| **Co$^{2+}$** | 4.23 | 2036 | 0.732 |
| **Pb$^{2+}$** | 4.01 | 1572 | 0.945 |
| **Al$^{3+}$** | 4.75 | 4715 | 0.541 |
| **NO$_3^-$** | 3.35 | 325 | 1.902 |
| **Cl$^-$** | 3.32 | 367 | 2.032 |

**Note**: The hydration enthalpies, hydration radii and the bulk diffusion coefficients of the ions listed above are taken from **Supplementary Ref.**[1], **Ref.**[2] and **Ref.**[3], respectively.

**Supplementary Table 2: Sequential permeation results for $K^+/Pb^{2+}$ pair tests across one $MoS_2$-COOH membrane**

| Test order | $K^+$ flux ratio (vs 1st $K^+$) | $Pb^{2+}$ flux ratio (vs 1st $K^+$) | Osmotic pressure ratio (vs 1st $K^+$) | Interlayer spacing (Å) |
|---|---|---|---|---|
| 1 | 1 | / | 1 | 11.1 |
| 2 | 0.973 | / | 1 | 11.1 |
| 3 | / | 1.328 | 1.5 | 11.4 |
| 4 | 0.818 | / | 1 | 11.4 |
| 5 | / | 1.443 | 1.5 | 11.4 |
| 6 | 2.556 | 1.246 | 2.5 | 11.4 |
| 7 | 2.691 | 1.437 | 2.5 | 11.4 |
| 8 | 1.74604 | 0.04375 | 1.075 | 11.4 |
| 9 | 0.83098 | / | 1 | 11.4 |
| 10 | 1.70535 | 0.00258 | 1.015 | 11.4 |
| 11 | 2.31309 | 0.13612 | 1.15 | 11.4 |
| 12 | 2.82796 | 0.40162 | 1.375 | 11.4 |
| 13 | 2.46634 | 0.63339 | 1.75 | 11.4 |
| 14 | 0.86316 | / | 1 | 11.4 |
| 15 | 2.66578 | 0.32244 | 1.3 | 11.4 |
| 16 | 2.80766 | 0.42746 | 1.375 | 11.4 |
| 17 | 2.62094 | 0.52055 | 1.5 | 11.4 |

**Note**: The osmotic pressure is calculated according to the van't Hoff law. We assume the van't Hoff factor is 1 in all cases, so the osmotic pressure ratio is the ratio of total ion molarity (cation and anion combined).

**Supplementary Table 3: 12-6-4 parameters set for the ions in conjunction with the OPC3 water model.**

| | $R_{min,M}/2$ (Å) | $\varepsilon_M (kcal/mol)$ | $C_M (kcal/mol \cdot Å^4)$ |
|---|---|---|---|
| $K^+$ | 1.760 | 0.18150763 | 16 |
| Pb²⁺ | 1.681 | 0.12564307 | 47 |
| Co²⁺ | 1.443 | 0.02387506 | 182 |
| Cl⁻ | 2.165 | 0.53403341 | -47 |

Note: $K^+$, $Co^{2+}$ and $Cl^-$ parameters are obtained from **Supplementary Ref.** [4] and [5], while $Pb^{2+}$ parameter is parametrized in the present work.

**Supplementary Table 4: Target and calculated values of the HFE, IOD, and CN in the first solvation shell for $Pb^{2+}$ ions.**

| | HFE (kcal/mol) | | IOD (Å) | | CN | |
|---|---|---|---|---|---|---|
| | Exp. [a] | Cal. | Exp. [b] | Cal. | Exp. [b] | Cal. |
| Pb²⁺ | -340.6 | -340.9 | 2.53 | 2.53 | 6 | 6.9 |

[a] The HFE values are from previous work[6].

[b] The IOD and CN values are extracted from the EXAFS fitting in our previous work (**Ref.33** of main text).

**Supplementary Table 5: Summary of $K^+$/$Pb^{2+}$ permeation test simulations using all-atom model developed in this study.**

| Test order | MoS$_2$ Fun. state | Number of particles | Ion number in Feed | Force (KJ/mol/nm) | Pressure (Mpa) | Ion number in Membrane | Ion number in Permeate | Simulation time (ns) |
|---|---|---|---|---|---|---|---|---|
| 1 | All-protonated | 45531 | 12 K$^+$ / 12 Cl$^-$ | 400 | 21.6 | 0 | | Run1: 54 ns<br>Run2: 52 ns<br>Run3: 52 ns |
| 2 | Half-deprotonated | 45885 | 30 Pb$^{2+}$ / 24 Cl$^-$ | 600 | 31.4 | 0 | | Run1: 48 ns<br>Run2: 46 ns<br>Run3: 58 ns |
| 3 | | 45585 | 12 K$^+$ / 12 Cl$^-$ | 400 | 21.6 | 18 Pb$^{2+}$ | | Run1: 90 ns<br>Run2: 90 ns<br>Run3: 100 ns |
| 4 | | 45588 | 12 K$^+$ / 1 Pb$^{2+}$ / 14 Cl$^-$ | 400 | 21.6 | 18 Pb$^{2+}$ | | Run1: 96 ns<br>Run2: 100 ns<br>Run3: 105 ns |
| 5 | | 45594 | 12 K$^+$ / 3 Pb$^{2+}$ / 18 Cl$^-$ | 550 | 29.7 | 18 Pb$^{2+}$ | 0 | Run1: 76 ns<br>Run2: 70 ns<br>Run3: 82 ns |
| 6 | | 45597 | 12 K$^+$ / 4 Pb$^{2+}$ / 20 Cl$^-$ | 600 | 32.4 | 18 Pb$^{2+}$ | | Run1: 56 ns<br>Run2: 50 ns<br>Run3: 54 ns |
| 7 | | 45603 | 12 K$^+$ / 6 Pb$^{2+}$ / 24 Cl$^-$ | 700 | 37.8 | 18 Pb$^{2+}$ | | Run1: 32 ns<br>Run2: 34 ns<br>Run3: 34 ns |
| 8 | | 45609 | 12 K$^+$ / 8 Pb$^{2+}$ / 28 Cl$^-$ | 800 | 43.2 | 18 Pb$^{2+}$ | | Run1: 28 ns<br>Run2: 28 ns<br>Run3: 26 ns |
| 9 | | 45589 | 14 K$^+$ / 14 Cl$^-$ | 400 | 21.6 | 18 Pb$^{2+}$ | | Run1: 105 ns<br>Run2: 98 ns<br>Run3: 105 ns |

**Supplementary Table 6: Sequential permeation results of $Ba^{2+}/Pb^{2+}$ pair tests across one $MoS_2$-COOH membrane.**

| Test order | $Ba^{2+}$ flux ratio (vs $1^{st}$ $Ba^{2+}$) | $Pb^{2+}$ flux ratio (vs $1^{st}$ $Ba^{2+}$) | Osmotic pressure ratio (vs $1^{st}$ $Ba^{2+}$) | Interlayer spacing (Å) |
|---|---|---|---|---|
| 1 | 1 | / | 1 | 13.1 |
| 2 | / | 0.69612 | 1 | 11.4 |
| 3 | 0.71345 | / | 1 | 11.4 |
| 4 | / | 0.75142 | 1 | 11.4 |
| 5 | 0.69509 | 0.01851 | 1.01 | 11.4 |
| 6 | 0.58378 | 0.06578 | 1.05 | 11.4 |
| 7 | 0.52689 | 0.11849 | 1.1 | 11.4 |
| 8 | 0.53796 | 0.29633 | 1.25 | 11.4 |
| 9 | 0.43502 | 0.44067 | 1.5 | 11.4 |
| 10 | 0.3918 | 0.76508 | 2 | 11.4 |
| 11 | 0.72439 | / | 1 | 11.4 |

**Supplementary Table 7: Sequential permeation results of $Co^{2+}$/$Pb^{2+}$ pair tests across one $MoS_2$-COOH membrane.**

| Test order | $Co^{2+}$ flux ratio (vs 1st $Co^{2+}$) | $Pb^{2+}$ flux ratio (vs 1st $Co^{2+}$) | Osmotic pressure ratio (vs 1st $Co^{2+}$) | Interlayer spacing (Å) |
|---|---|---|---|---|
| 1 | 1 | / | 1 | 12.7 |
| 2 | / | 0.7403 | 1 | 11.4 |
| 3 | 0.74152 | / | 1 | 11.4 |
| 4 | / | 0.84269 | 1 | 11.4 |
| 5 | 0.65374 | 0.02162 | 1.01 | 11.4 |
| 6 | 0.51756 | 0.09079 | 1.05 | 11.4 |
| 7 | 0.43316 | 0.16371 | 1.1 | 11.4 |
| 8 | 0.34101 | 0.23594 | 1.2 | 11.4 |
| 9 | 0.31405 | 0.48591 | 1.5 | 11.4 |
| 10 | 0.25108 | 0.92989 | 2 | 11.4 |
| 11 | 0.69428 | / | 1 | 11.4 |

**Supplementary Table 8: Summary of $Co^{2+}/Pb^{2+}$ permeation test simulations using all-atom model developed in this study.**

| Test order | $MoS_2$ Fun. state | Number of particles | Ion number in Feed | Force (KJ/mol/nm) | Pressure (Mpa) | Ion number in Membrane | Ion number in Permeate | Simulation time (ns) |
|---|---|---|---|---|---|---|---|---|
| 1 | All-protonated | 45615 | 12 $Co^{2+}$ / 24 $Cl^-$ | 400 | 21.6 | 0 | 0 | Run1: 73 ns<br>Run2: 64 ns<br>Run3: 68 ns |
| 2 | Half-deprotonated | 45597 | 12 $Co^{2+}$ / 24 $Cl^-$ | 400 | 21.6 | 18 $Pb^{2+}$ | | Run1: 160 ns<br>Run2: 117 ns<br>Run3: 155 ns |
| 3 | | 45600 | 12 $Co^{2+}$ / 1 $Pb^{2+}$ / 26 $Cl^-$ | 400 | 21.6 | 18 $Pb^{2+}$ | | Run1: 160 ns<br>Run2: 142 ns<br>Run3: 164 ns |
| 4 | | 45606 | 12 $Co^{2+}$ / 3 $Pb^{2+}$ / 30 $Cl^-$ | 500 | 27.0 | 18 $Pb^{2+}$ | | Run1: 108 ns<br>Run2: 99 ns<br>Run3: 100 ns |
| 5 | | 45609 | 12 $Co^{2+}$ / 4 $Pb^{2+}$ / 32 $Cl^-$ | 534 | 28.84 | 18 $Pb^{2+}$ | | Run1: 90 ns<br>Run2: 82 ns<br>Run3: 81 ns |
| 6 | | 45615 | 12 $Co^{2+}$ / 6 $Pb^{2+}$ / 36 $Cl^-$ | 600 | 32.4 | 18 $Pb^{2+}$ | | Run1: 107 ns<br>Run2: 92 ns<br>Run3: 108 ns |

**Supplementary References:**